\documentclass[aps,prb,twocolumn,showpacs,floatfix]{revtex4-2}
\usepackage{amssymb}
\usepackage{graphicx,epsfig,psfrag,subfigure,amsopn,epstopdf}
\usepackage{amsmath}
\usepackage{color}
\usepackage{bm}
\begin{document}
\newcommand\sg{\subfigure}
\newcommand\non{\nonumber}
\newcommand\noi{\noindent}
\newcommand\new{\newpage}
\newcommand\ig{\includegraphics}
\newcommand\De{\Delta}
\newcommand{\tb}[1]{\textcolor{magenta}{#1}}
\title{Relativistic particles in super-periodic potentials: exploring graphene and fractal systems}
\author{Sudhanshu Shekhar}
\author{Bhabani Prasad Mandal}
\author{Anirban Dutta}
\email{anirband@bhu.ac.in}
\affiliation{Department of Physics, Banaras Hindu University, Varanasi-221005, India}
%
\begin{abstract}
In this article, we employ the transfer matrix method to investigate relativistic particles in super-periodic potentials (SPPs) of arbitrary order $n \in I^{+}$. We calculate the reflection and transmission probabilities for spinless Klein particles encountering rectangular potential barriers with super-periodic repetition. It is found that spinless relativistic particles exhibit Klein tunneling and a significantly higher degree of reflection compared to their non-relativistic counterparts. Additionally, we analytically explore the behavior of experimentally realizable massless Dirac electrons as they encounter rectangular potential barriers with a super-periodic pattern in a monolayer of graphene. In this system, the transmission probability, conductance, and Fano factor are evaluated as functions of the number of barriers, the order of super-periodicity, and the angle of incidence. Our findings reveal that the transmission probability shows a series of resonances that depend on the number of barriers and the order of super-periodicity. We extend our analysis to specific cases within the Unified Cantor Potentials (UCPs)-$\gamma$ system ($\gamma$ is a scaling parameter greater than $1$), focusing on the General Cantor fractal system and the General Smith-Volterra-Cantor (GSVC) system. For the General Cantor fractal system, we calculate the tunneling probability, which reveals sharp transmission peaks and progressively thinner unit cell potentials as $G$ increases. In the GSVC system, we analyze the potential segment length and tunneling probability, observing nearly unity tunneling coefficients when $\gamma \approx 1$, as well as saturation behavior in transmission coefficients at higher stages $G$.
\end{abstract}
\date{\today}
\maketitle
\section{Introduction}
In recent times, the study of transport properties in graphene like materials subjected to external scalar potentails have gathered more attraction in research both theoretically \cite{novoselov2005two, novoselov2004electric, zhang2005experimental, park2008anisotropic, gumbs2014revealing, hui2024two, brey2009emerging, ohta2012evidence, li2024transport, ahmad2024transverse, zhang2005electric, berger2004ultrathin, wang2021global, azarova2014transport, xu2010electron, dakhlaoui2021quantum, pakdel2021transport, zhan2013transfer} and experimentally \cite{huard2007transport, williams2007quantum, oostinga2008gate, gorbachev2008conductance, liu2008fabrication, cheianov2006selective, abanin2007quantized}. Scattering phenomena arising from locally periodic potentials significantly affect a wide range of physical systems, including quantum particle dynamics and electron transport in materials \cite{park2008anisotropic, gumbs2014revealing, griffiths2001waves, griffiths1992scattering, lee1989one, hasan2020tunneling, xu2015transmission}. A precise understanding of these phenomena is essential for advancing our knowledge across various fields, including quantum transport, quantum Hall effect, \cite{hasegawa2013periodic, park2009landau, wang2015evidence, dean2013hofstadter, usov1988theory, kol1993fractional, tan1994localization}, and electronics \cite{capasso1990resonant, slater1949electrons}. In graphene, the linear dispersion at the Fermi points results in the presence of Dirac-like quasiparticles. At the six high-symmetry points in the hexagonal Brillouin zone, the energy bands touch the Fermi energy; however, only two of these points are distinct and exhibit different chiralities. This results in perfect transmission known as Klein tunneling, and graphene offers an experimental platform for testing the Klein paradox \cite{beenakker2008colloquium, allain2011klein, stander2009evidence, huard2007transport, williams2007quantum, oostinga2008gate, gorbachev2008conductance, liu2008fabrication, cheianov2006selective}. Moreover, the ability to modify and control transport properties with external scalar and vector potentials has drawn considerable interest from the scientific community because of its potential implications for numerous technological applications \cite{williams2007quantum, oostinga2008gate, gorbachev2008conductance, liu2008fabrication, cheianov2006selective, abanin2007quantized}. \\
In this work, we investigate scattering behaviors of relativistic particles in pesence of SPPs. A SPPs \cite{hasan2018super} refers to a periodic potentials in which additional periodic modulations or variations are superimposed on top of the main periodic structure. Essentially, it involves introducing secondary periodicity into a system that already exhibits periodic behavior. SPPs are of significant interest because they can lead to novel physical phenomena \cite{hasan2018super, umar2023quantum, umar2024polyadic, singh2023quantum, narayan2023tunneling}. They can influence the electronic structure of materials, affect transport properties, and lead to the emergence of new electronic states that are not present in purely periodic systems. Therefore, studying SPPs is crucial for understanding the behavior of materials in diverse physical conditions and for exploring potential applications in areas such as semiconductor devices, photonic crystals, and more.\\
We have used the transfer matrix method to develop a theoretical framework for studying Klein tunneling in SPPs of any order. This framework is applicable to both spinless quantum particles described by the Klein-Gordon equation and massless Dirac electrons governed by the Dirac equation. Our investigation focuses on two intriguing scenarios. First, we examine the behavior of relativistic quantum particles encountering super-periodic rectangular potential barriers, comparing the dynamics of spinless Klein particles with their non-relativistic equivalents in these settings. Our findings indicate a significant difference in behavior, with Klein particles demonstrating a greater propensity for reflection. Notably, they also show a finite probability of tunneling through infinitely large super-periodic barriers, which contradicts classical predictions. In the second part of our study, we analyze massless Dirac electrons in monolayer graphene as they interact with super-periodic electrostatic barriers. Our objective is to uncover the complex relationship between the number of barriers, the order of super-periodicity, and the angle of incidence, particularly in terms of transmission probability, conductance, and the Fano factor. We reveal an intriguing pattern of resonances in the transmission probability and explain the behaviors of conductance and the Fano factor in relation to super-periodicity.
In this article, we explore the intriguing phenomenon of Klein tunneling in SPPs and apply these insights to practical advancements. By examining the relationship between relativistic quantum mechanics and periodic structures, we aim to investigate these phenomena across various materials that host relativistic particles, thereby deepening our understanding of transmission of relativistic particle under SPPs. This research could pave the way for the development of next-generation materials that harness the relativistic nature of the quasiparticles.\\
We organize the paper as follows: In Sec.~\ref{sec:II}, we provide a brief overview of the the general theory of wave propagation in a locally periodic and super-periodic media. Following this, we derive the expressions for transmission and reflection coefficient for Klein particle in presence of SPPs by extending the formulation developed in \cite{hasan2018super}. We describe the Klein paradox and the formation of resonance bands in our system using the expression for the reflection coefficient. Next in Sec.~\ref{sec:III} We discuss electron tunneling in graphene in presence of periodic and SPPs. We derive the expression for reflection and transmission coefficient and explain the resonance bands, Klein tunneling and conductance in shot noise experiment. We extend our formalism to fractal potentials in Sec.~\ref{sec:IV} and derive the expression for tunneling coefficient. Finally, in Sec.~\ref{sec:V}, we summarize our findings and conclude. We present the detailed calculations necessary to determine the reflection and transmission coefficients in a SPPs, in appendix.\\
\section{Relativistic particle in SPPs}
\label{sec:II}
\begin{figure}[htb]
\centering
\includegraphics[width=\columnwidth]{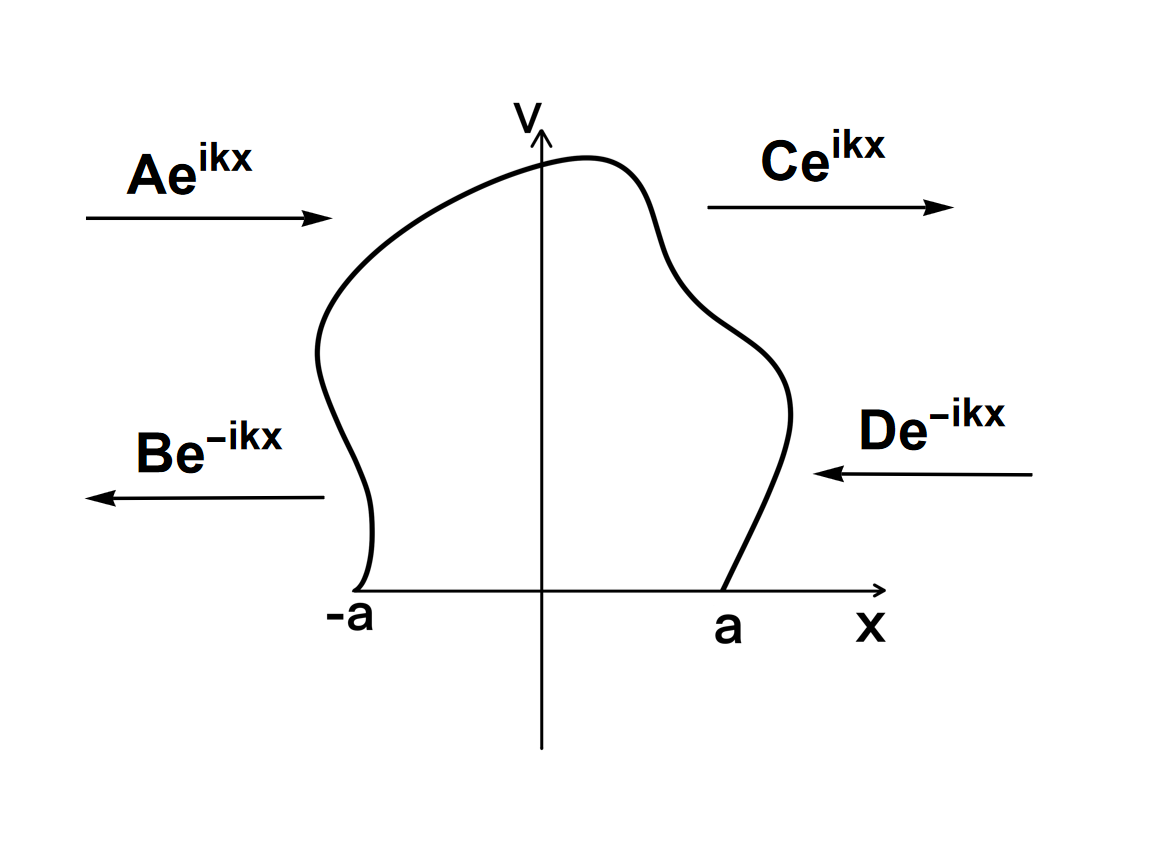}
\caption{Schematic diagram of particle scattering from a arbitrary potential.}
\label{fig:1}
\end{figure}

\begin{figure}[htb]
\centering
\includegraphics[width=\columnwidth]{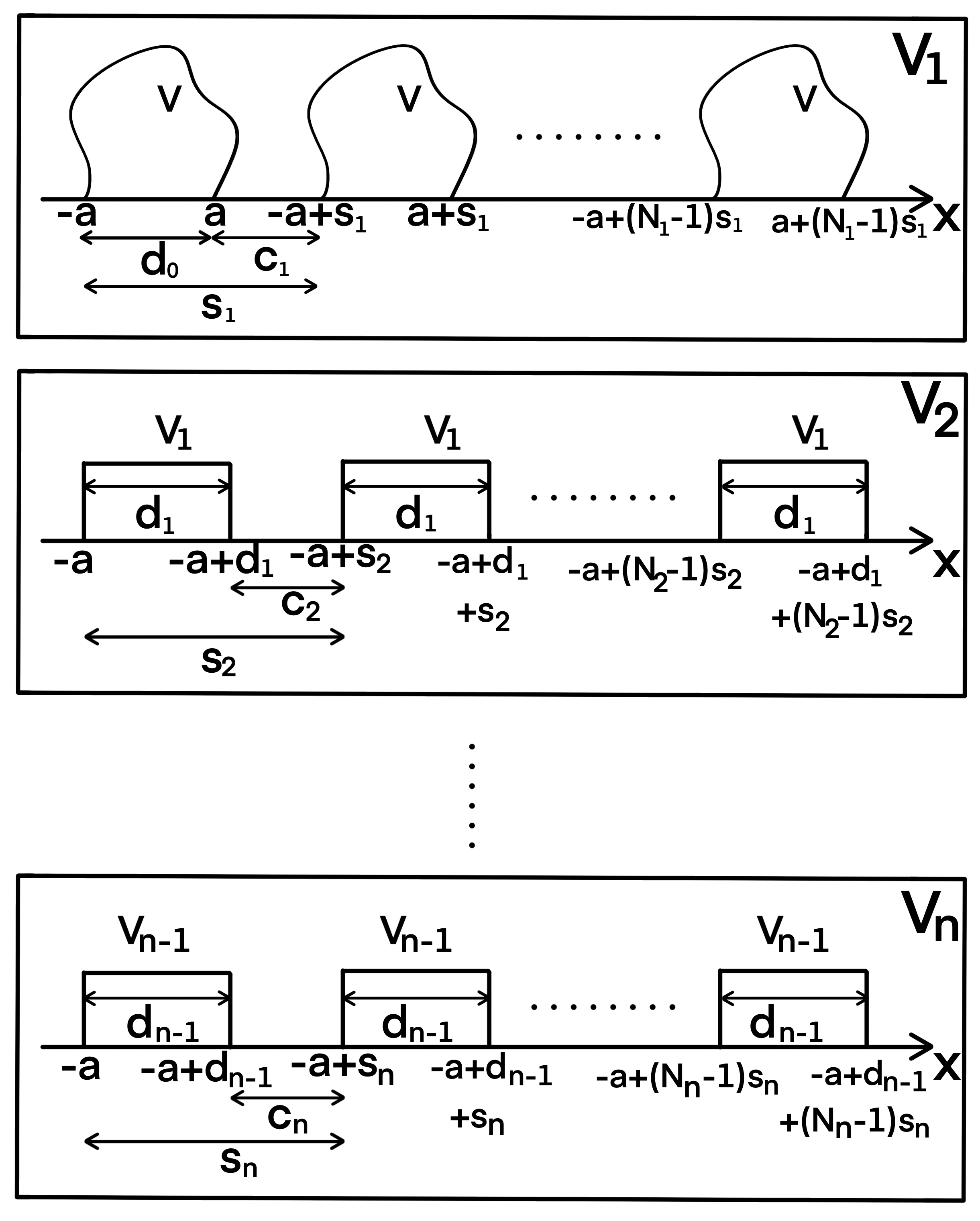}
\caption{Schematic diagram of SPPs.}
\label{fig:2}
\end{figure}

In this section, we introduce SPPs in detail and discuss the behaviour of relativistic particle in presence of those.  Let us consider the locally periodic potentials illustrated in FIG. \ref{fig:2}., where the potential $V(x)$ of a single unit cell is repeated periodically $N_{1}$ times over a distance $s_{1}$ along the $x$-direction (with $s_{1}=d_{0}+c_{1}$, $d_{0}=2a$, and $c_{1}>0$. If we know the transfer matrix for one unit cell potential, then we can obtain the transfer matrix for the entire periodic system. We briefly discuss the essential points of the theory here. Let us consider a potential $V(x)$ that spans an interval of length $d_{0}=2a$, $-a<x<a$, which serves as the unit cell of a periodic structure that repeats itself $N_{1}$ times, with cells separated by a distance $s_{1}=d_{0}+c_{1},~c_{1}>0$, as shown in the top FIG. \ref{fig:2}. 
This periodic potentials has a total span of $d_{1} = N_{1}d_{0} + (N_{1}-1)c_{1}$ and is treated as the whole potential $V_{1}$. In case of super-periodicity, the potential $V_{1}$ is repeated $N_{2}$ times with a separation of $s_{2}=d_{1}+c_{2},~c_{2}>0$. The total span of the SPPs $V_{2}$ is given by $d_{2} = N_{2}d_{1} + (N_{2}-1)c_{2}$. We define the SPPs of order $n$, after repeating this procedure $n$ times, which is spanned in intervals,
\begin{eqnarray}
d_{n} = N_{n}d_{n-1} + (N_{n}-1)c_{n};~~ n\in I^{+}
\end{eqnarray}
It is worth noting that M.~Hasan \textit{et al.}~\cite{hasan2018super} derived closed-form expressions for each element of the transfer matrix in the context of super-periodic repetition of potentials. In this work, we have generalized their results to the case of relativistic particles. Moreover, this calculation does not depend on the specific shape of the potential of the single `unit cell' as long as one can find the transfer matrix of that `unit cell'. \\
We start with a brief description of the scattering of a relativistic particle in the presence of a arbitrary potential. We examine a potential $V(x)$ within the interval $(-a, a)$, with a relativistic particle approaching from the left, as shown in FIG.~\ref{fig:1}. Here, the potential is specifically a barrier potential rather than an arbitrary one. However, the calculations are more general and applicable to any form of potential, which is discussed in detail in the appendix. We focus on spinless quantum particles that obey the one-dimensional time-independent Klein-Gordon equation.
\subsection{Klein particle in the SPPs}
{The time-independent Klein-Gordon equation for a particles moving in one dimension is given by 
\begin{eqnarray}
\frac{d^{2}\psi}{dx^{2}}+\left[\frac{(E-V_{0})^{2}-m^{2}c^{4}}{\hbar^{2}c^{2}}\right]\psi=0,
\label{K.G.}
\end{eqnarray}
where $\psi$ is the wave function of the particle having mass $m$ and energy $E$. $c$ is the speed of light. 
Considering the reflection and transmission across the potential barrier, the general solution of Eq.($\ref{K.G.}$) in each of the three regions can be expressed as:
\begin{eqnarray}
\psi(x) = \begin{cases} 
Ae^{iqx}+Be^{-iqx} &\mbox{for } x<-a\\
Ee^{iq'x}+Fe^{-iq'x} & \mbox{for } -a<x<a\\
Ce^{iqx}+De^{-iqx} & \mbox{for }  x>a 
\end{cases}
\label{wave function Klein}
\end{eqnarray}
Where $q=\sqrt{\frac{E^{2}-m^{2}c^{4}}{\hbar^2c^2}}$ and $q'=\sqrt{\frac{(E-V_{0})^{2}-m^{2}c^{4}}{\hbar^2c^2}}$ are the wave vectors outside and inside the barrier respectively. Using the boundary condition of continuity of the wave function and its derivative at $x=-a$ and $x=a$, we find the transfer matrix as defined in the appendix. The elements of the transfer matrix for the Klein particle are:
\begin{eqnarray}
m_{11}&=&\left(\cos{(2q'a)}-i\epsilon_{+}\sin{(2q'a)}\right)e^{2iqa}\label{m11kg}=m_{22}^{*}\\
m_{12}&=&i\epsilon_{-}\sin{(2q'a)}\label{m12kg}=m_{21}^{*}
\end{eqnarray}
where, $\epsilon_{\pm}=\frac{1}{2}\left(\alpha\pm{1/\alpha}\right)\label{24}$ and, $\alpha={q}/{q'}$. This transfer matrix of the rectangular barrier is building block to calculate the transmission and reflection probabilities for the locally periodic and SPPs of any order. The transmission probabilities for Klein particle in presence of a locally periodic rectangular barrier with a finite $N_1$ times repetations of one unit cell is given by
\begin{eqnarray}
T(N_{1},q)&=&\frac{1}{1+[\epsilon_{-}\sin(2q'a)U_{N_{1}-1}(\xi_{1})]^{2}}\label{28}
\end{eqnarray}
Following the same prescription as discussed in appendix it is also possible to get analytical formula for the transmission probabilities for the super-periodic rectangular barrier of order-$2$, using the equations (\ref{super periodic}) and (\ref{22a}) respectively,
\begin{eqnarray}
T(N_{1},N_{2},q)&=&\frac{1}{1+[\epsilon_{-}\sin(2q'a)\prod_{r=1}^{2}(U_{N_{r}-1}(\xi_{r}))]^{2}}
\label{29}\nonumber\\
\end{eqnarray}
where, 
\begin{eqnarray}
\xi_{1}&=&\text{Re}(m_{11})\cos(qs_{1})+\text{Im}(m_{11})\sin(qs_{1})\\
\xi_{2}&=&\text{Re}((m_{11})_{1})\cos(qs_{2})+\text{Im}((m_{11})_{1})\sin(qs_{2})
\end{eqnarray}
\begin{equation}
(m_{11})_{1}=m_{11}U_{N_{1}-1}(\xi_{1})e^{iq(N_{1}s_{1}-s_{1})}-U_{N_{1}-2}(\xi_{1})e^{iqN_{1}s_{1}}
\end{equation}
and
\begin{equation}
s_{1}=2a+c_{1}; ~ s_{2}=2aN_{1} + (N_{1}-1)c_{1}+c_{2}
{\label{s1}}
\end{equation}
One can straightforwardly compute the reflection probabilities for locally periodic and SPPs from Eqs. (\ref{28}), (\ref{29}) and given by  $R(N_{1},q)=1-T(N_{1},q)$ and $R(N_{1},N_2,q)=1-T(N_{1},N_2,q)$ respectively. These expressions are one of the important results we derive in this paper. In a simiar way it is also possible to extend the formalism further to get analytical expression for reflection and transmission probability for -periodic rectangular barrier for any arbitrary order $n$.} \\
To bemchmark our calculation, we derive the non-relativistic limits of the reflection and transmission coefficients for SPPs and compare them with the known results found in the literature\cite{hasan2018super}. In the non-relativistic regime the total energy $E=mc^{2}+E_{nr}$ is comparable with $E_{nr}$, and $V_{0}$ and small compare to $mc^{2}$.
\begin{equation}
\begin{gathered}
E=mc^{2}+E_{nr}\approx mc^{2}\\
E+mc^{2}\approx 2mc^{2}; ~ E-mc^{2}=E_{nr}
\end{gathered} 
\end{equation}
where $E_{nr}$ is energy of the system in non-relativistic case. Relativistic momentums $q$ also get modified to non-relativistic momentum $k$ and given by, 
\begin{eqnarray}
q\to k=\sqrt{\frac{2mE_{nr}}{\hbar^2}};~~~~~q'\to k'=\sqrt{\frac{2m(E_{nr}-V_{0})}{\hbar^2}}\nonumber\\
\end{eqnarray}
In the non-relativistic limit, the transmission probabilities for a periodic rectangular potentials simplify to:
\begin{eqnarray}
T(N_{1},k)&=&\frac{1}{1+[\epsilon_{-}\sin(2k'a)U_{N_{1}-1}((\xi_{1})_{nr})]^{2}}
\label{31a}\\
T(N_{1},N_{2},k)&=&\frac{1}{1+[\epsilon_{-}\sin(2k'a)\prod_{r=1}^{2}(U_{N_{r}-1}(\xi_{r})_{nr})]^{2}}
\label{31b}\nonumber\\
\end{eqnarray}
where, 
\begin{eqnarray}
(\xi_{1})_{nr}&=&\text{Re}(m_{11})\cos(ks_{1})+\text{Im}(m_{11})\sin(ks_{1})\\
(\xi_{2})_{nr}&=&\text{Re}((m_{11})_{1})\cos(ks_{2})+\text{Im}((m_{11})_{1})\sin(ks_{2})\nonumber\\
\end{eqnarray}
\begin{equation}
(m_{11})_{1}=m_{11}U_{N_{1}-1}(\xi_{1})e^{ik(N_{1}s_{1}-s_{1})}-U_{N_{1}-2}(\xi_{1})e^{ikN_{1}s_{1}}
\end{equation}
Since the reflection and transmission coefficients add up to one, the reflection coefficients $R(N_{1},k)$ and $R(N_{1},N_{2},k)$ can be easily calculated using Eqs.(\ref{31a}), (\ref{31b}) and in the non-relativistic limit our results match with the result in the literature \cite{hasan2018super}.\\
\begin{figure}[htb]
\begin{center}
\sg[~$N_{1}=2$]{\ig[height=3.1cm]{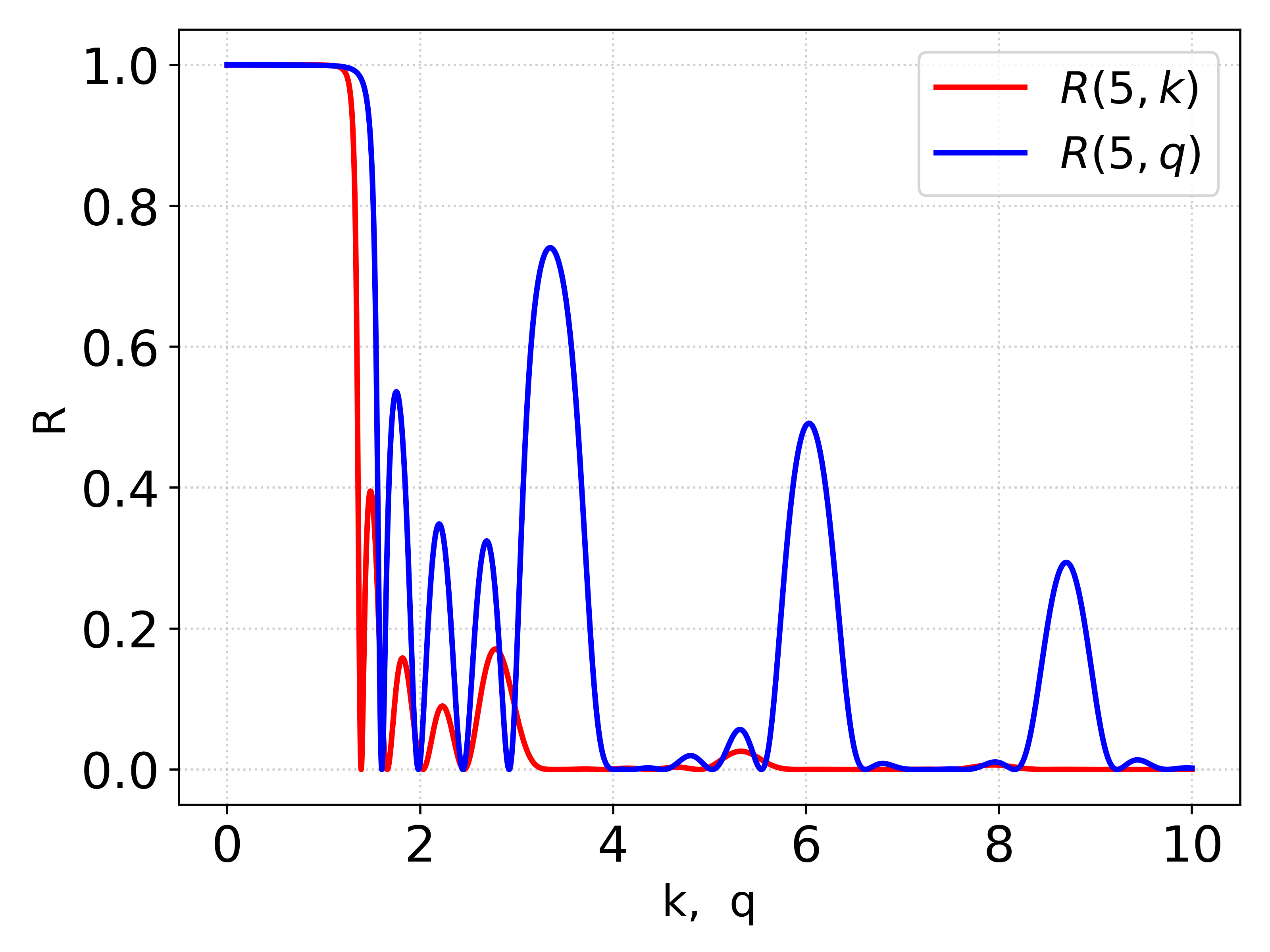}\label{fig:3a}}
\sg[~$N_{1}=2,~N_{2}=2$]{\ig[height=3.1cm]{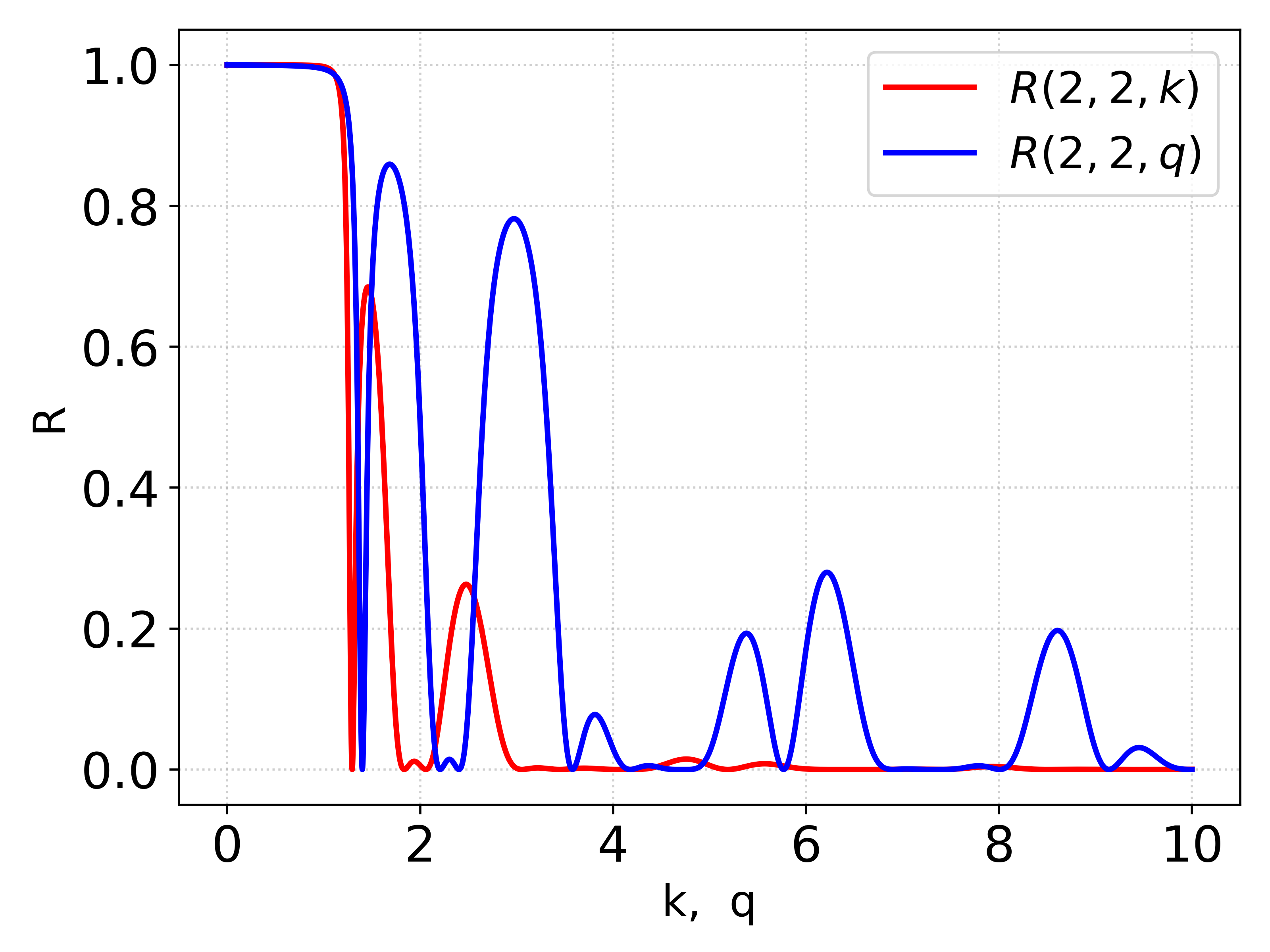}\label{fig:3b}}\\
\sg[~Klein Tunneling]{\ig[height=3.1cm]{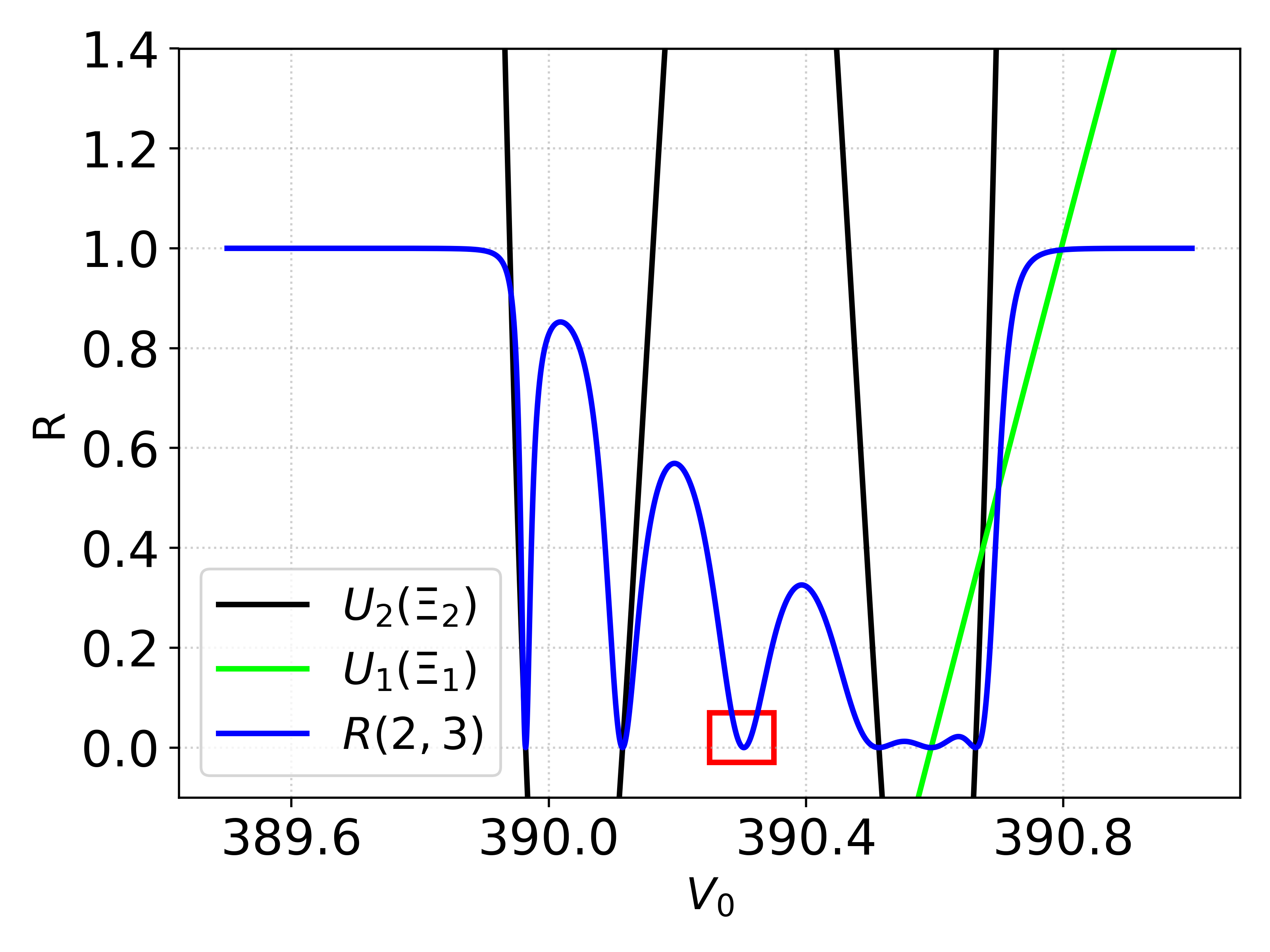}\label{fig:3c}}
\sg[~$\text{Bloch Phase}$]{\ig[height=3.1cm]{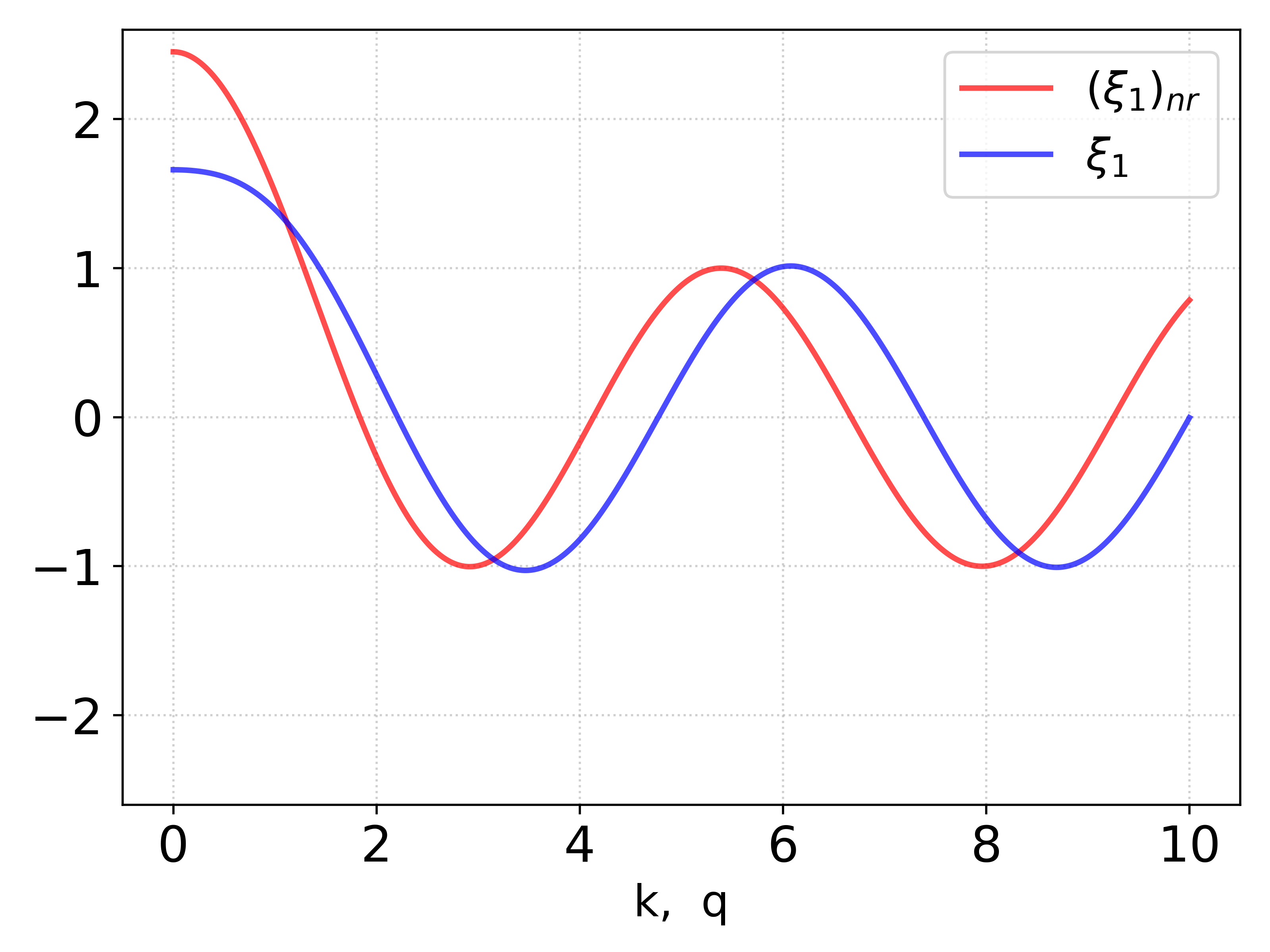}\label{fig:3d}}
\end{center}
\caption{{The behaviour of reflection probability for both relativistic (blue curve) and non-relativistic (red curve) particle in presence of periodic FIG. \ref{fig:3a} and SPPs of order-$2$ FIG. \ref{fig:3b}. Here $a=0.5,$ $V_{0}=1, m=1,c_{1}=0.2, \text{ and } c_{2}=0.6$ for plots. In the third plot FIG. \ref{fig:3c}, the reflection coefficient approaches zero at large potential values where cosine of the Bloch phases also vanish. However, an additional dip in the reflection coefficient is observed when the condition $2V_{0}a=\beta\pi$ is satisfied. The last plot shows the cosine of the Bloch phase in the periodic system for both relativistic (blue curve) and non-relativistic (red curve) cases.}} 
\label{fig:3} 
\end{figure}
%
We have plotted the reflection probabilities for both relativistic (blue) and nonrelitivistic (red) case in FIG. \ref{fig:3}. In certain energy ranges, the reflection probability in the relativistic case approaches zero, similar to the behavior observed in the non-relativistic case. However, in these energy ranges, the relativistic wave experiences a higher degree of reflection compared to its non-relativistic counterpart as shown in FIG. \ref{fig:3a} and FIG. \ref{fig:3b}. The Chebyshev polynomials (CPs) of the second kind can be expressed in terms of sinusoidal functions as follows: 
\begin{equation}
U_{N_{1}}(\xi_{1}) = \frac{\sin((N_{1}+1)\gamma)}{\sin(\gamma)}\label{11}
\end{equation}
Let $N_1$ be a non-negative integer, and let $\xi_1$ take real values in the interval $[-1, 1]$. The parameter $\gamma = \cos^{-1}(\xi_1)$ is commonly referred to as the Bloch phase. As $N_1 \to \infty$, total reflection occurs when $\xi_1$ lies outside the interval $[-1, 1]$. This is because the CP diverges to infinity outside this interval as $N_1 \to \infty$.
The cosine of the Bloch phase is illustrated in FIG.~\ref{fig:3d} for both relativistic and non-relativistic cases. The relativistic case demonstrates similar transmission resonances.
\subsubsection{Klein Tunneling}
In this subsection we demonostrate Klein-tunneling when relativistic particle encounter SPPs. In the limit of $V_{0}\to\infty$ the transmission probability given by Eq.(\ref{28}) for the periodic potentials or SPPs of order-$1$ have the form:
\begin{equation}
T(N_{1},q)\to\frac{4}{4+[V_{0}\sin(2V_{0}a)U_{N_{1}-1}(\Xi_{1})]^{2}}\label{at infinite potential}
\end{equation}
While, the transmission probability for the SPPs of arbitrary order-$n$ as $V_{0}\to\infty$ is:
\begin{eqnarray}
T(N_{1},N_{2},\cdots,q)&\to&\frac{4}{4+[V_{0}\sin(2V_{0}a)\prod_{r=1}^{n}(U_{N_{r}-1}(\Xi_{r}))]^{2}}\nonumber\\
\label{nth order}
\end{eqnarray}
where $\Xi$ is the argument of the CP at $V_{0}\to\infty$. This indicates that when $2V_{0}a = \beta\pi$, with $\beta$ an integer, or when the CP is zero, the transmission coefficient is one and the reflection coefficient is zero, for any order. When the transmission coefficient equals one due to the CP being zero, it is referred to as transmission resonance. On the other hand, when $2V_{0}a = \beta\pi$ and the transmission coefficient equals one, this phenomenon is known as Klein tunneling as demonstrated in FIG. \ref{fig:3c} and is exclusively attributed to the negative energy solution.

The graph in FIG. \ref{fig:3c} exhibits six distinct dips, five of these originate from the zeros of the CPs $U_{1}(\Xi_{1})$ and $U_{2}(\Xi_{2})$, while the sixth dip, enclosed in the red box, results from the condition $2V_{0}a = \beta\pi$, with $\beta \approx 124 $. It has the same origin as the Klein paradox and our calculation explicitly show the presence of Klein tunneling in presence of SPPs of arbitrary order. 
\section{Electron tunneling through super-periodic electrostatic barriers in graphene}
\label{sec:III}
In the previous section, we have established the formalism for a relativistic particle in a SPPs. In this section, we now focus on a more experimentally realizable system of quantum particles, which are governed by the relativistic Dirac equation. 
We consider a monolayer graphene in the $xy$ plane. On the top of the monolayer, we apply an electrostatic potential barrier  that has a rectangular shape and is infinite along the $y$ axis
\begin{equation}
V(x)=\begin{cases}
V_{0},~~-a< x< a\\0,~~\text{otherwise}
\end{cases}
\end{equation}
The experimental realization of this type of barrier potential has already been achieved \cite{huard2007transport, williams2007quantum, oostinga2008gate, gorbachev2008conductance, liu2008fabrication, cheianov2006selective, abanin2007quantized}. We assume that the electron incidence at an angle $\phi$ to the $x$-axis as shown in FIG. \ref{fig:4}.  The motion of the electrons is described by the massless Dirac equation
\begin{equation}
(-i\hbar v_{f}\sigma_{x,y}\nabla+V(x))\psi(x,y)=E\psi(x,y).
\label{dirac equation}
\end{equation}
Where $v_{f}\approx10^{6}~\text{ms}^{-1}$ is the Fermi velocity, $\sigma_{x,y}$ are the Pauli matrices and $E$ is the energy. The Dirac spinor $\psi_{1}$ and $\psi_{2}$ can be written as \cite{katsnelson2006chiral}:
\begin{figure}[t]
\centering
\includegraphics[width=0.8\columnwidth]{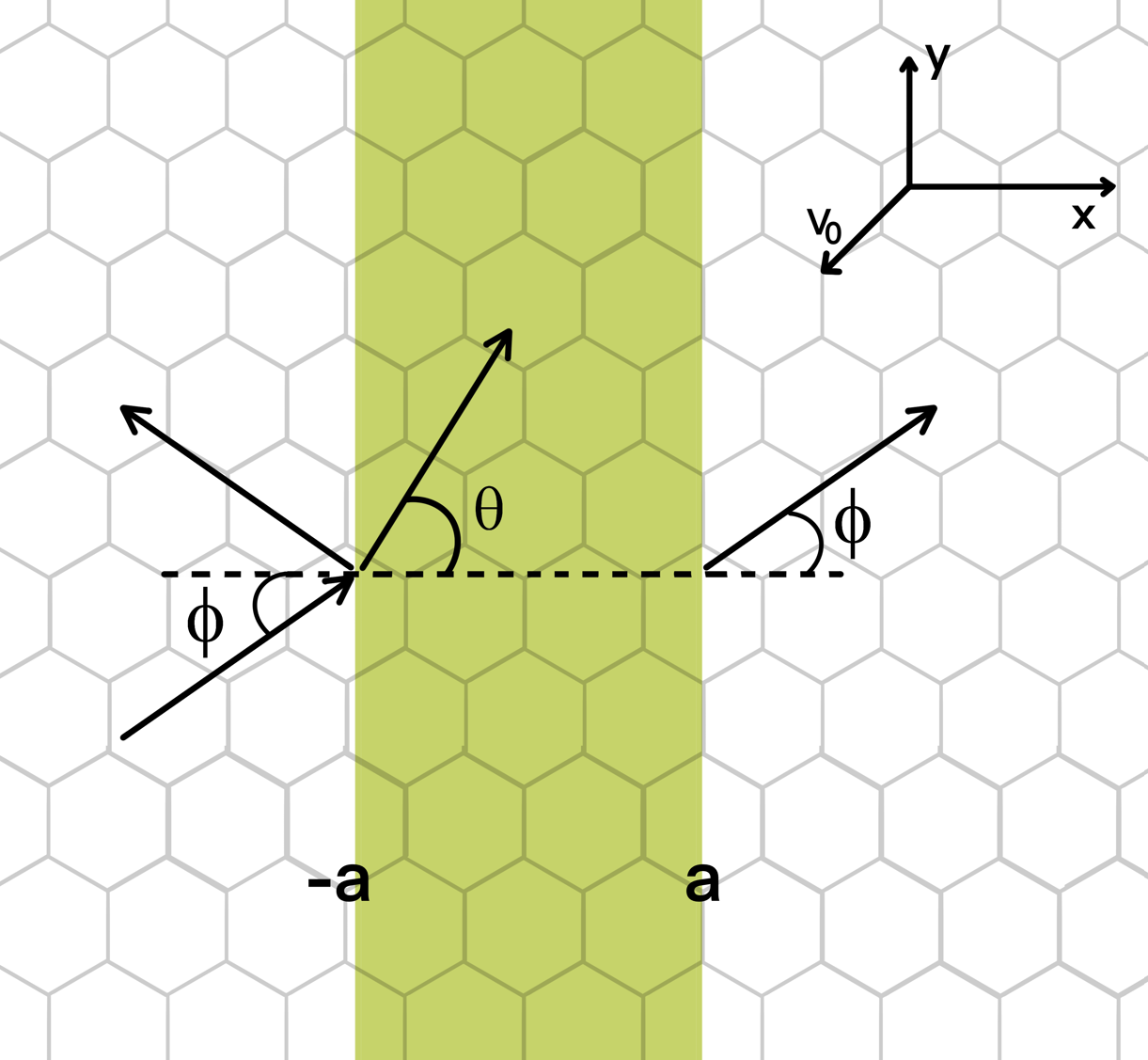}
\caption{Schematic diagram of a graphene monolayer under an external rectangular electrostatic potential.}
\label{fig:4}
\end{figure}
\begin{subequations}
\begin{align}
\psi_{1}(x,y)&=\begin{cases}
(Ae^{iq_{x}x}+Be^{-iq_{x}x})e^{iq_{y}y}, &\mbox{for } x<-a\\
(Ee^{i{q_{x}}^{\prime}x}+Fe^{-i{q_{x}}^{\prime}x})e^{iq_{y}y}, &\mbox{for } -a<x<a\\
(Ce^{iq_{x}x}+De^{-iq_{x}x})e^{iq_{y}y}, &\mbox{for } x>a
\end{cases}
\end{align}
\end{subequations}
\begin{subequations}
\begin{align}
\psi_{2}(x,y)&=\begin{cases}
s(Ae^{iq_{x}x+i\phi}-Be^{-iq_{x}x-i\phi})e^{iq_{y}y},~\mbox{for } x<-a\\
s'(Ee^{i{q_{x}}^{\prime}x+i\theta}-Fe^{-i{q_{x}}^{\prime}x-i\theta})e^{iq_{y}y},\\
\hspace{4cm}\mbox{for } -a<x<a\\
s(Ce^{iq_{x}x+i\phi}-De^{-iq_{x}x-i\phi})e^{iq_{y}y},~\mbox{for } x>a
\end{cases}
\end{align}
\end{subequations}
where $q_{f}=\frac{E}{\hbar v_{f}}$ is Fermi wavevector, $q_{x}=q_{f}\cos{\phi}$, $q_{y}=q_{f}\sin{\phi}$ are $x$ and $y$-components of the Fermi wavevector outside of the barrier respectively, ${q_{x}}^{\prime}=\sqrt{\frac{(E-V_{0})^{2}}{\hbar^{2}v_{f}^{2}}-q_{y}^{2}}$, $\theta=\tan^{-1}\left(\frac{q_{y}}{{q_{x}}^{\prime}}\right)$, $s=\text{sgn}(E)$, and $s'=\text{sgn}(E-V_{0})$.
The transfer matrix can be obtained by using the continuity of the wave function at the boundaries $x=-a$ and $x=a$. The elements of the transfer matrix are:
\begin{eqnarray}
m_{11}&=&\frac{1}{ss'}e^{2iq_{x}a} (ss'\cos (2{q_{x}}^{\prime}a)-i\sin(2{q_{x}}^{\prime}a)\nonumber\\
&&\quad\times(\sec(\theta) \sec(\phi)-ss'\tan(\theta )\tan (\phi)))\label{m11}=m_{22}^{*}\nonumber\\ \\
m_{12}&=&\frac{1}{2ss'}\sec(\theta)e^{-i(2q_{x}a+\phi)}\sin(2{q_{x}}^{\prime}a)\nonumber\\
&&\quad\times(-2ss'\sin(\theta) \sec(\phi)+2\tan(\phi))\label{m12}=m_{21}^{*}
\end{eqnarray}
\onecolumngrid
\begin{widetext}
\begin{figure}[htb]
\begin{center}
\sg[~$N_{1}=1$]{\ig[height=6cm]{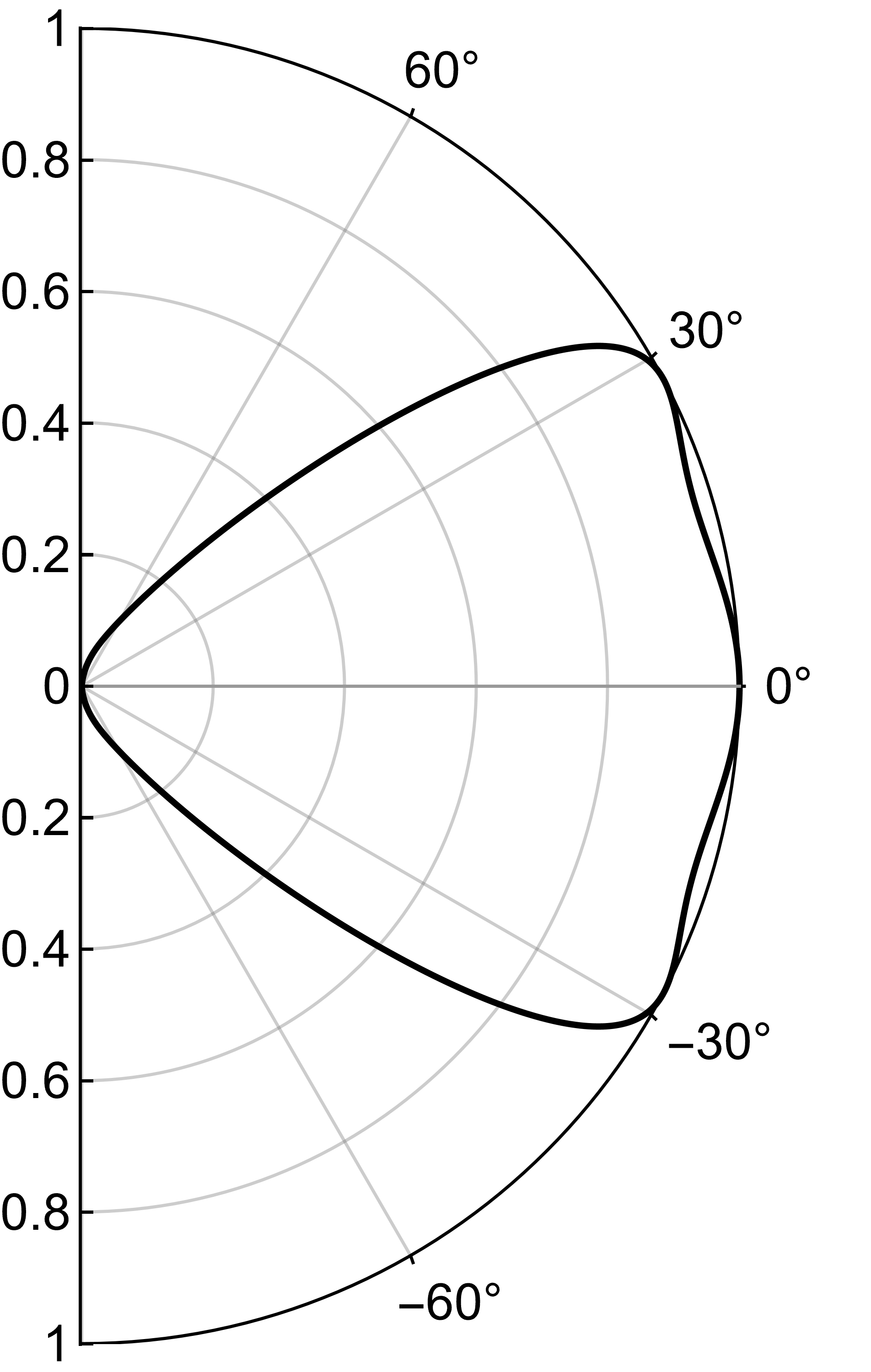}\label{fig:5a}}
\hspace{0.5cm}
\sg[~$N_{1}=5$]{\ig[height=6cm]{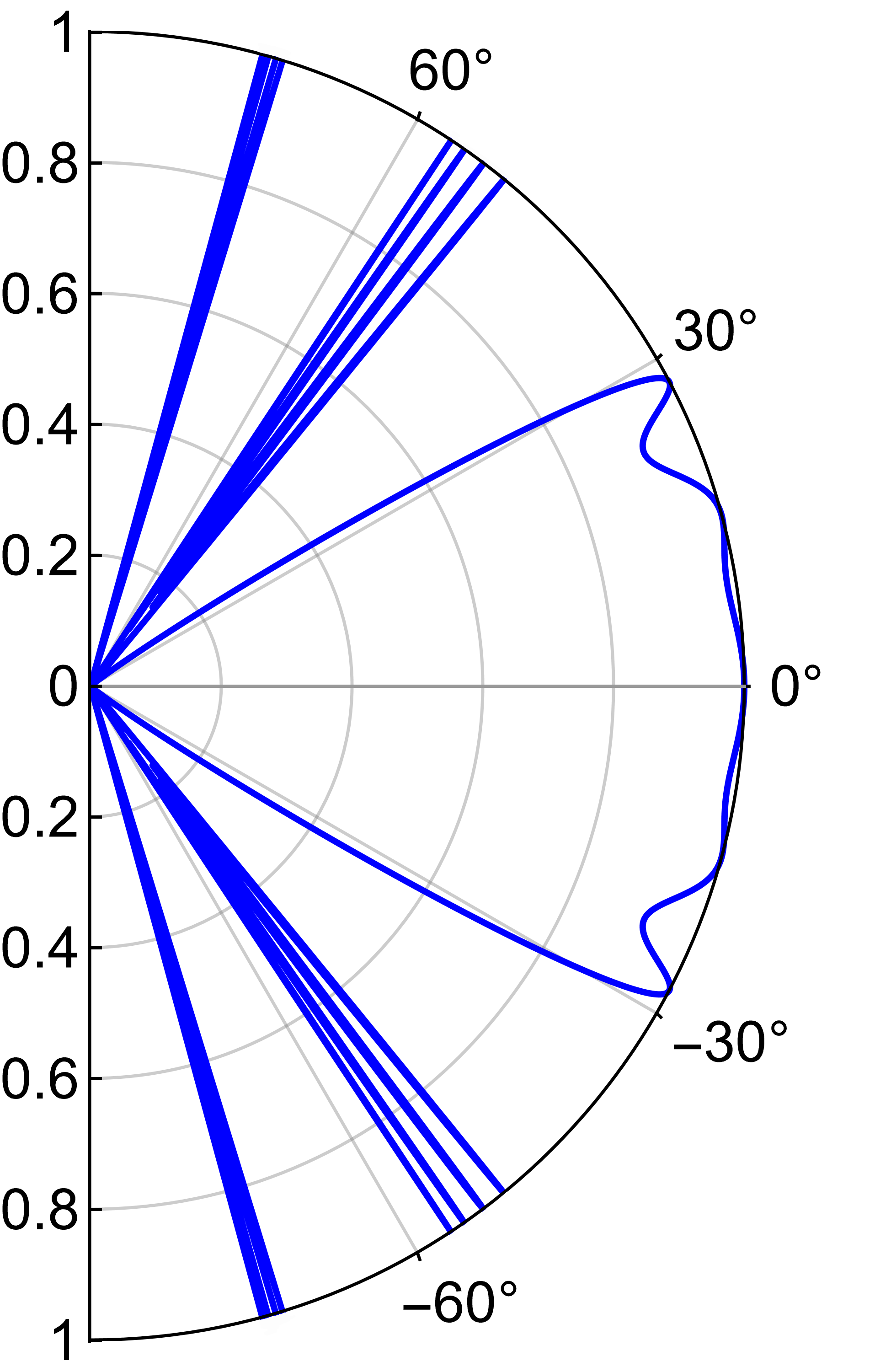}\label{fig:5b}}
\hspace{0.5cm}
\sg[~$N_{1}=8$]{\ig[height=5.7cm]{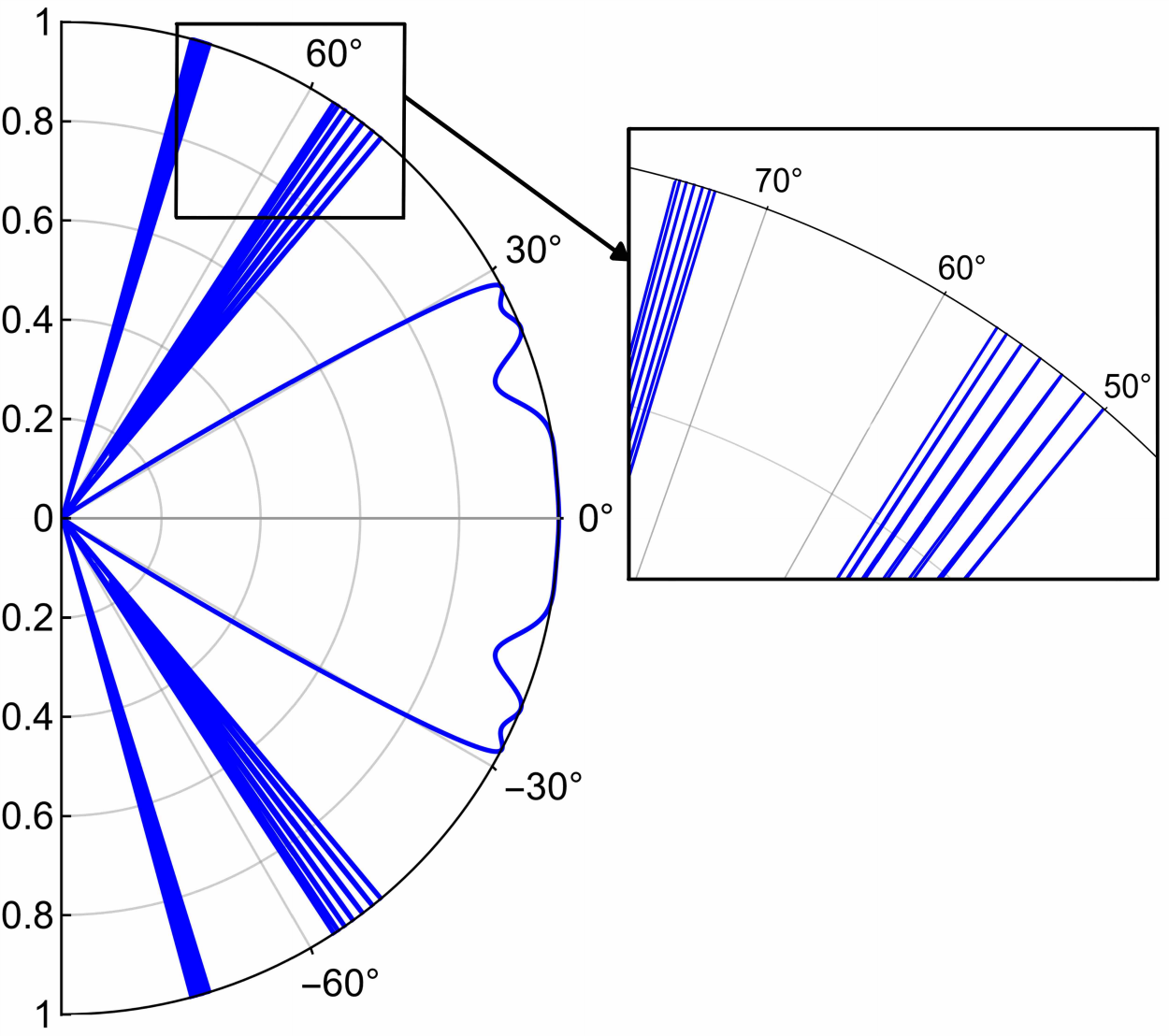}\label{fig:5c}}
\end{center}
\caption[n1 values]{Polar plot of transmission probability for a single (Black) and periodic (Blue) electrostatic potential(s). For these plots, the energy of the electron is $E = 80\text{ meV}$. The widths of the barriers and the separation between successive barriers are $2a = 100\text{ nm}$ and $c_{1} = 90\text{ nm}$, respectively. The Fermi velocity is $v_{f} = 10^{6}\text{ m/s}$ and the barrier height is $V_{0} = 250\text{ meV}$.}
\label{fig:5} 
\end{figure}
\end{widetext}
The matrix element $m_{12}$ is utilized to compute the transmission probability for both periodic and super-periodic electrostatic potentials of any order. The transmission probabilities, $T(N_{1}, \phi)$ and $T(N_{1}, N_{2}, \phi)$, corresponding to a periodic barrier and a super-periodic barrier of order-2 at a given energy $E$, are expressed as follows:
\begin{eqnarray}
T(N_{1},\phi)&=&\frac{1}{1+\left[\Delta(N_{1},\phi)U_{N_{1}-1}(\xi_{1})\right]^{2}} \label{periodic transmission}\\
T(N_{1},N_{2},\phi)&=&\frac{1}{1+\left[\Delta(N_{1},\phi)U_{N_{1}-1}(\xi_{1})U_{N_{2}-1}(\xi_{2})\right]^{2}}\nonumber\\
\label{spp transmission}
\end{eqnarray}
%

%
%
where, 
\begin{eqnarray}
\Delta(N_{1},\phi)&=&\sin(2{q_{x}}^{\prime}a)(ss'\tan(\theta )\sec(\phi )-\sec(\theta)\tan(\phi))\\
\xi_{1}&=&\text{Re}(m_{11})\cos(q_{x}s_{1})+\text{Im}(m_{11})\sin(q_{x}s_{1})\\
\xi_{2}&=&\text{Re}((m_{11})_{1})\cos(q_{x}s_{2})+\text{Im}((m_{11})_{1})\sin(q_{x}s_{2})\nonumber\\ \\
(m_{11})_{1}&=&m_{11}U_{N_{1}-1}(\xi_{1})e^{iq_{x}(N_{1}s_{1}-s_{1})}-U_{N_{1}-2}(\xi_{1})e^{iq_{x}N_{1}s_{1}}\nonumber\\
\label{}
\end{eqnarray}
And, the expression of the parameters $s_{1}$ and $s_{2}$ are the same as equations (\ref{s1}).
\onecolumngrid
\begin{widetext}
\begin{figure}[h]
\begin{center}
\sg[~$N_{1}=4,~N_{2}=2$]{\ig[height=6cm]{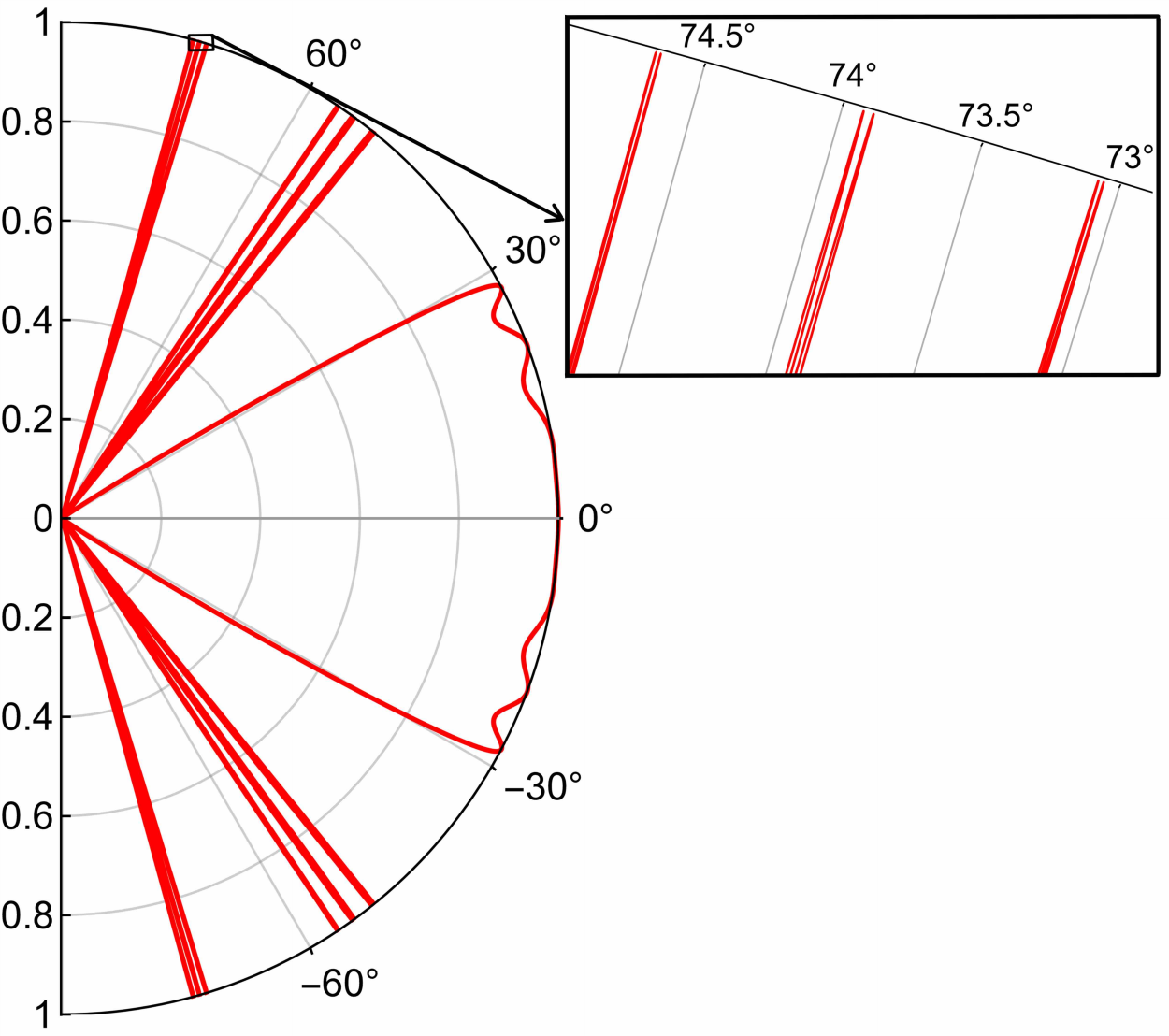}\label{fig:6a}}
\sg[~$N_{1}=4,~N_{2}=5$]{\ig[height=6cm]{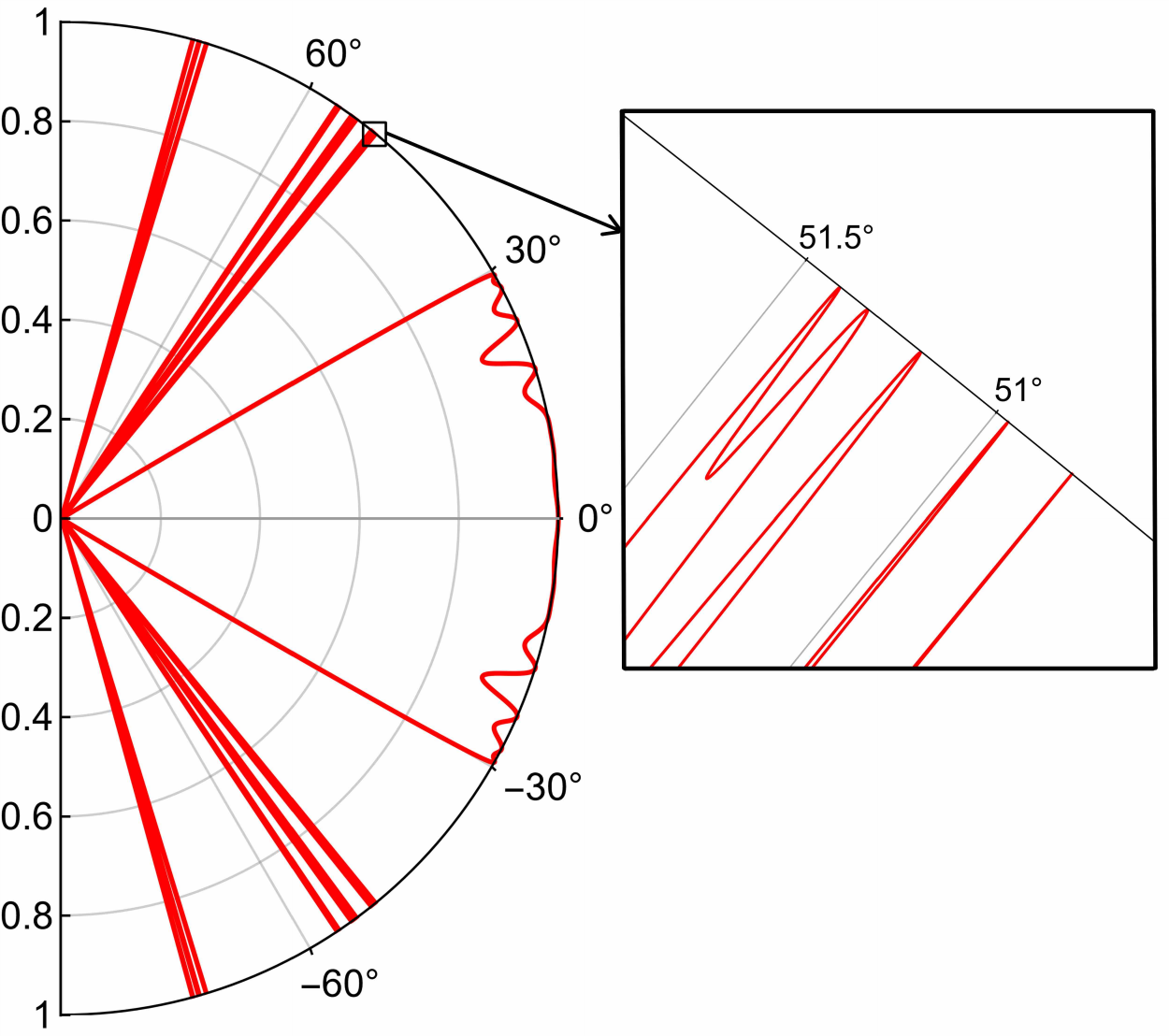}\label{fig:6b}}
\end{center}
\caption{Polar plot of transmission probability for the super-periodic electrostatic potentials of order-$2$. The number of barriers varies from $N_{2} = 2$ and $N_{2} = 5$ and the fix value of $N_{1}=4$. In this case, the separation between successive barriers is $c_{2} = 60\text{ nm}$, and all other parameters are the same as the FIG. \ref{fig:5}.}
\label{fig:6} 
\end{figure}
\end{widetext}
%
The effect of the number of electrostatic barriers on the transmission probabilities for periodic potentials and super-periodic electrostatic potentials is shown in FIG. \ref{fig:5} and FIG. \ref{fig:6}. We observed that for $\phi=0^{\circ}$, the transmission coefficient equals unity and does not depend on the number of electrostatic barriers for both periodic and super-periodic cases. This behavior confirms the Klein-tunneling effect, which states that the system is completely transparent for normal incidence, even for large barrier widths. Additionally, as $\phi$ increases, the peaks of the transmission coefficient become sharper. We also observe that the transmission curves are symmetric concerning the normal axis $(\phi = 0^\circ)$, regardless of the number of barriers. When the number of barriers increases $N_{1}>1$, the transmission probability exhibits four additional groups of peaks and each group has $N_{1}-1$ peaks. These peaks are available in a certain range of incidence angles.
Furthermore, we analyze the impact of the number of electrostatic barriers on the transmission coefficient for the super-periodic electrostatic potentials of order-$2$. As $\phi$ increases, the transmission coefficient peaks become sharper, resembling the behavior observed in the periodic potentials. However, for the SPPs of order-$2$, each peak of the periodic potentials contains $N_{2}$ resonance peaks, as illustrated in FIG.~\ref{fig:6}. Resolving these resonance peaks becomes increasingly challenging for higher values of $N_{2}$.\\
\begin{figure}[b]
	\sg[~]
	{\ig[height=2.5cm]{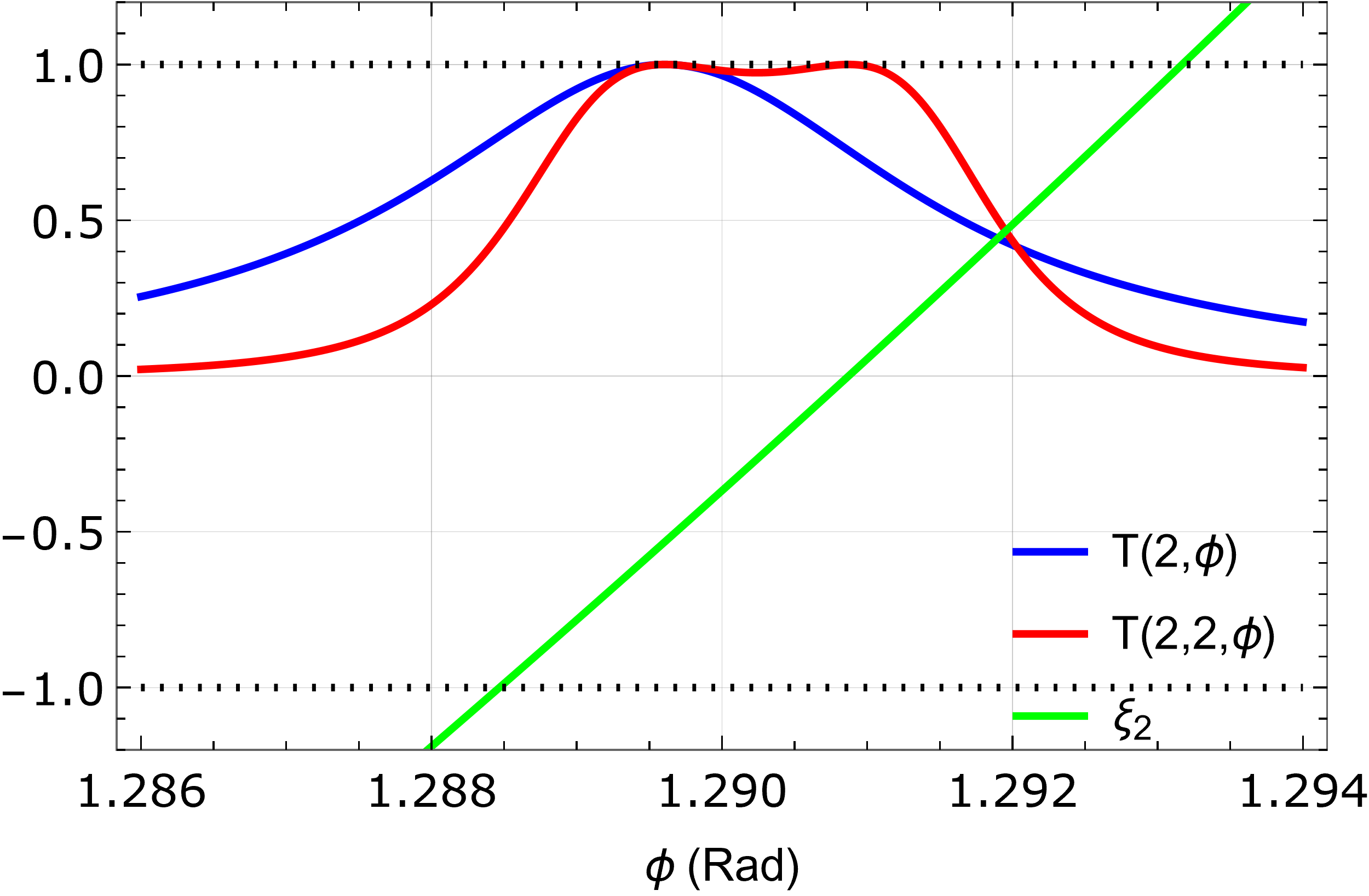}\label{fig:7a}}
	\sg[~]{\ig[height=2.5cm]{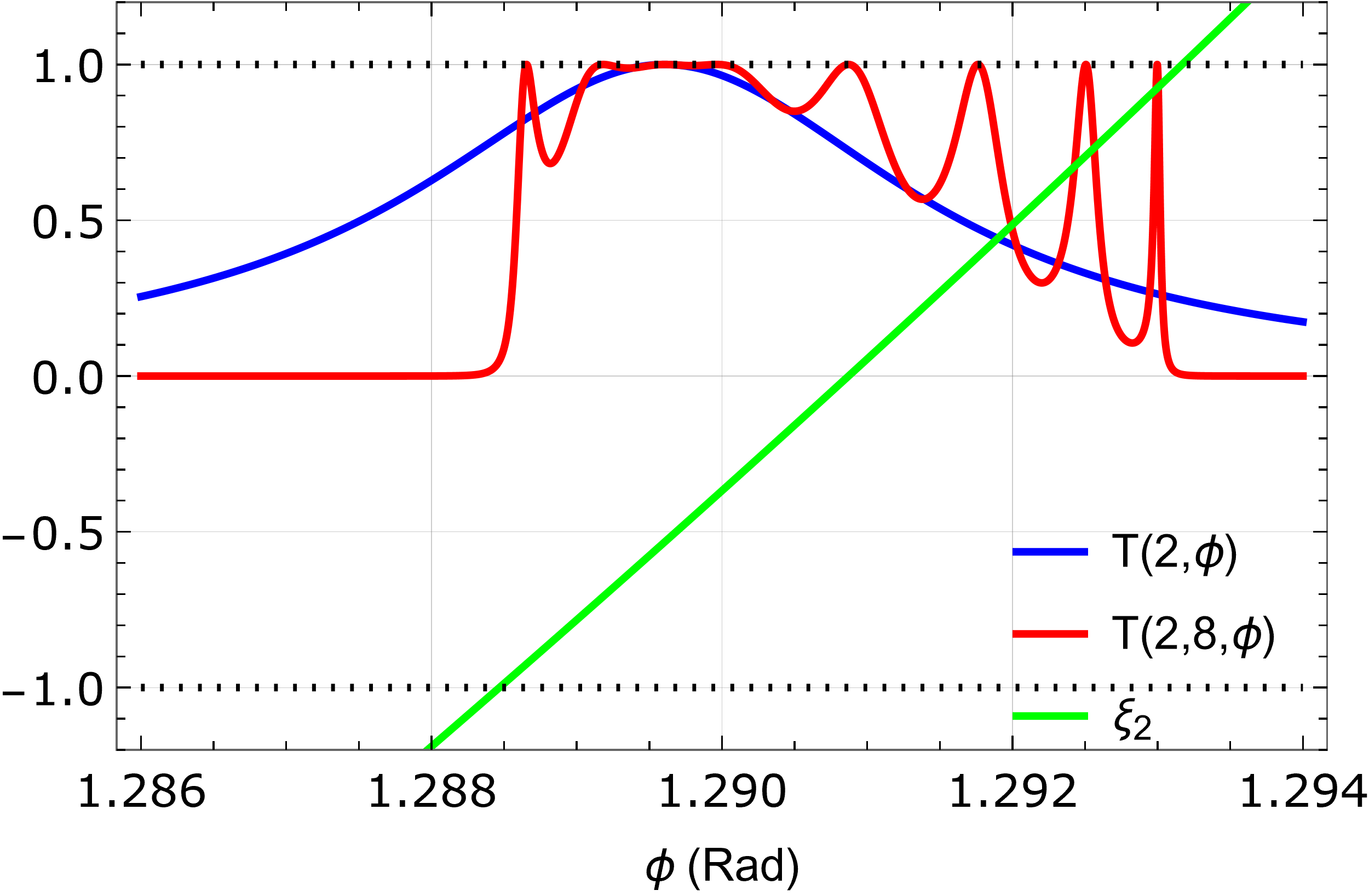}\label{fig:7b}}\\
	\sg[~]
	{\ig[height=2.5cm]{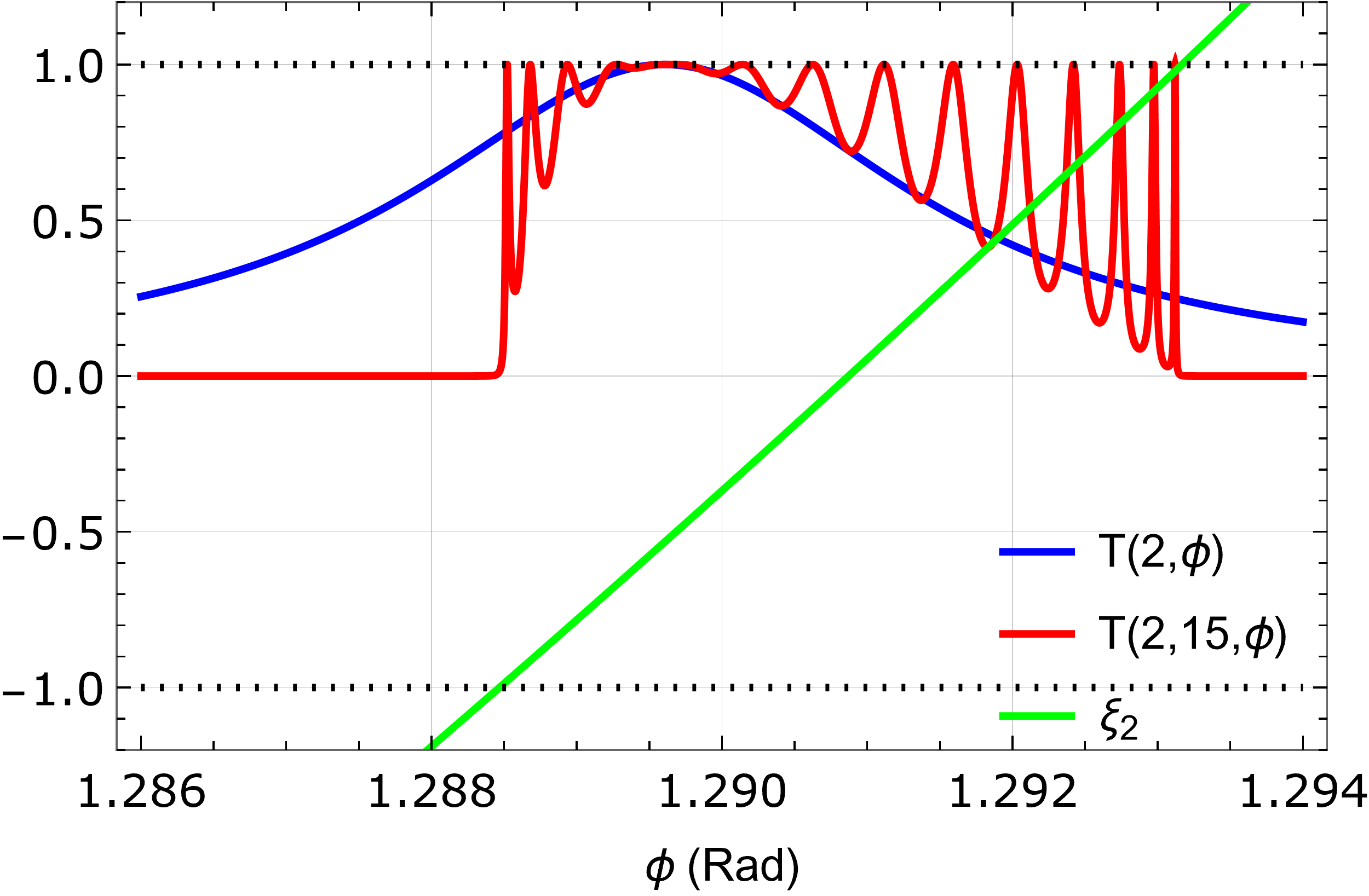}\label{fig:7c}}
	\sg[~]{\ig[height=2.5cm]{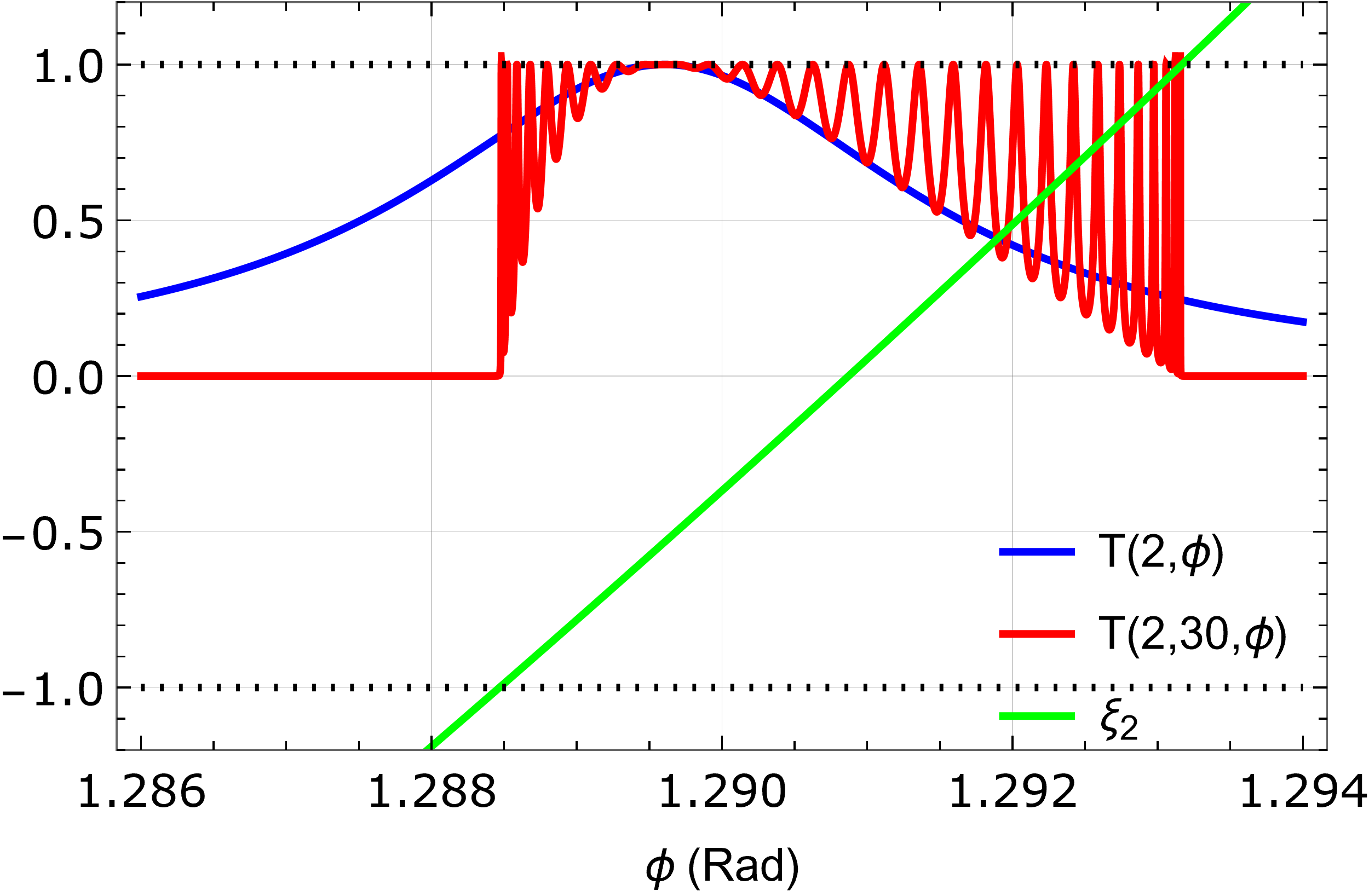}\label{fig:7d}}\\
	\caption{The dotted lines at $-1$ and +1 are included to illustrate that the resonance band is located within the range where $\xi_{2}\in[-1,1]$. The parameters for these plots are the same as the FIG. \ref{fig:6}.} 
	\label{fig:7} 
\end{figure}
\subsubsection{Resonance band:} In this subsection we explain the reason of $N_{2}$ number of resonance peaks observed in transmission coefficient. The CP $U_{N_{2}-1}(\xi_{2})$ in the equation (\ref{spp transmission}) determines the $N_{2}$ number of resonance peaks. A CP of degree $N_{2}-1$ has $N_{2}-1$ roots within the interval from $-1$ to $1$. Therefore, equation (\ref{spp transmission}) exhibits that there are $N_{2}-1$ resonance peaks when $\xi_{2}$ is between $-1$ and $1$, and one additional peak comes from the equation $U_{N_{1}-1}(\xi_{1})=0$. Therefore, each resonance band will contain a total of $(N_{2}-1)+1=N_{2}$ resonance peaks. We demonstrate this in the FIG. \ref{fig:7}, which shows the variation of transmission probability for different $N_{2}$, the resonance band contains $N_{2}$ number of resonance peaks. From these figures, it is also evident that in the neighborhood of transmission resonance of the corresponding periodic electrostatic potentials, the bloch phase $\xi_{2}$ lies between $-1$ to $1$. We observe that as the value of $N_{2}$ increases, it becomes more challenging to distinguish individual resonance peaks.
\onecolumngrid
\begin{widetext}
\begin{figure}[htb]
\begin{center}
\sg[~]{\ig[height=4cm]{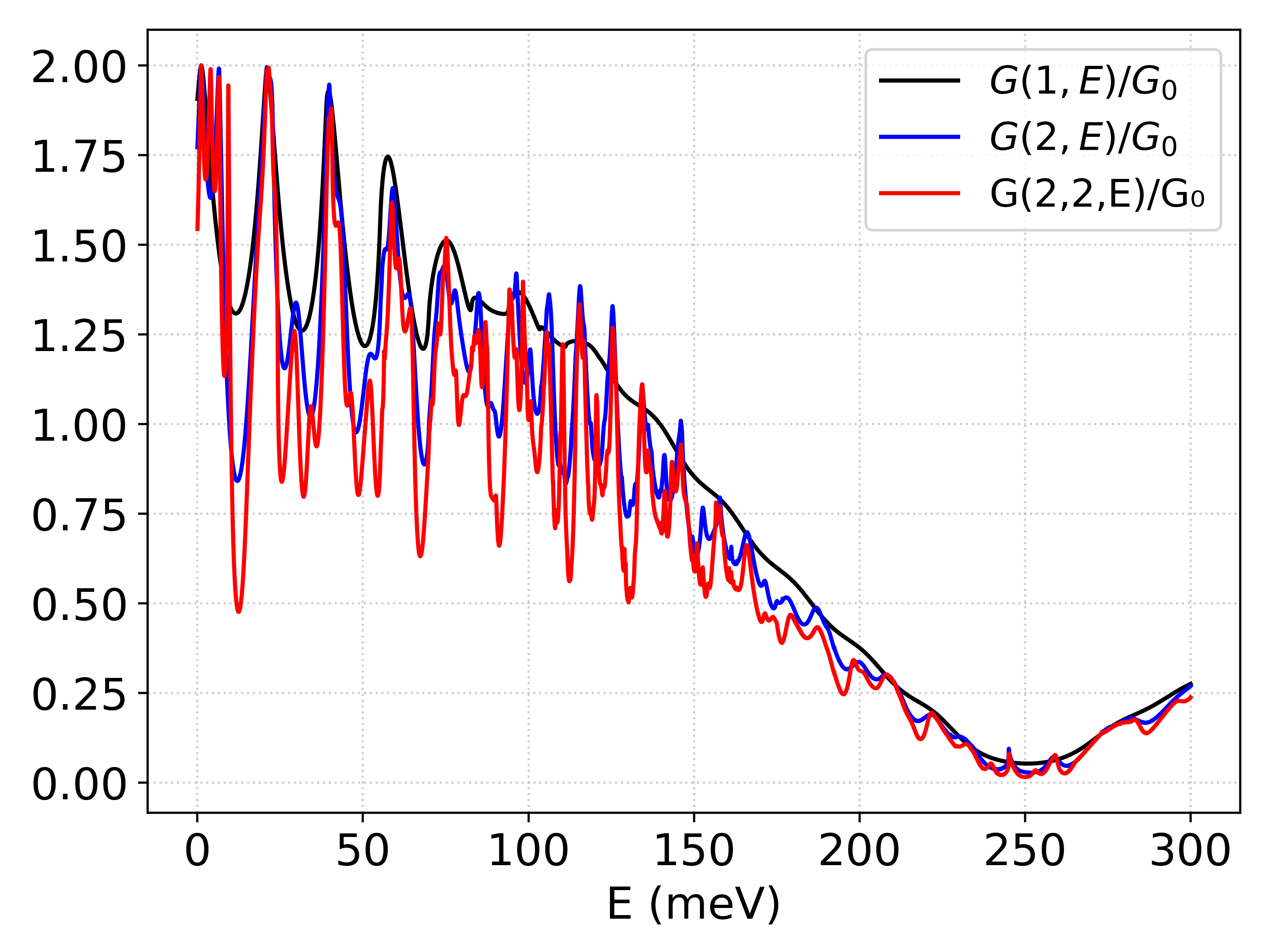}\label{fig:8a}}
\sg[~]{\ig[height=4cm]{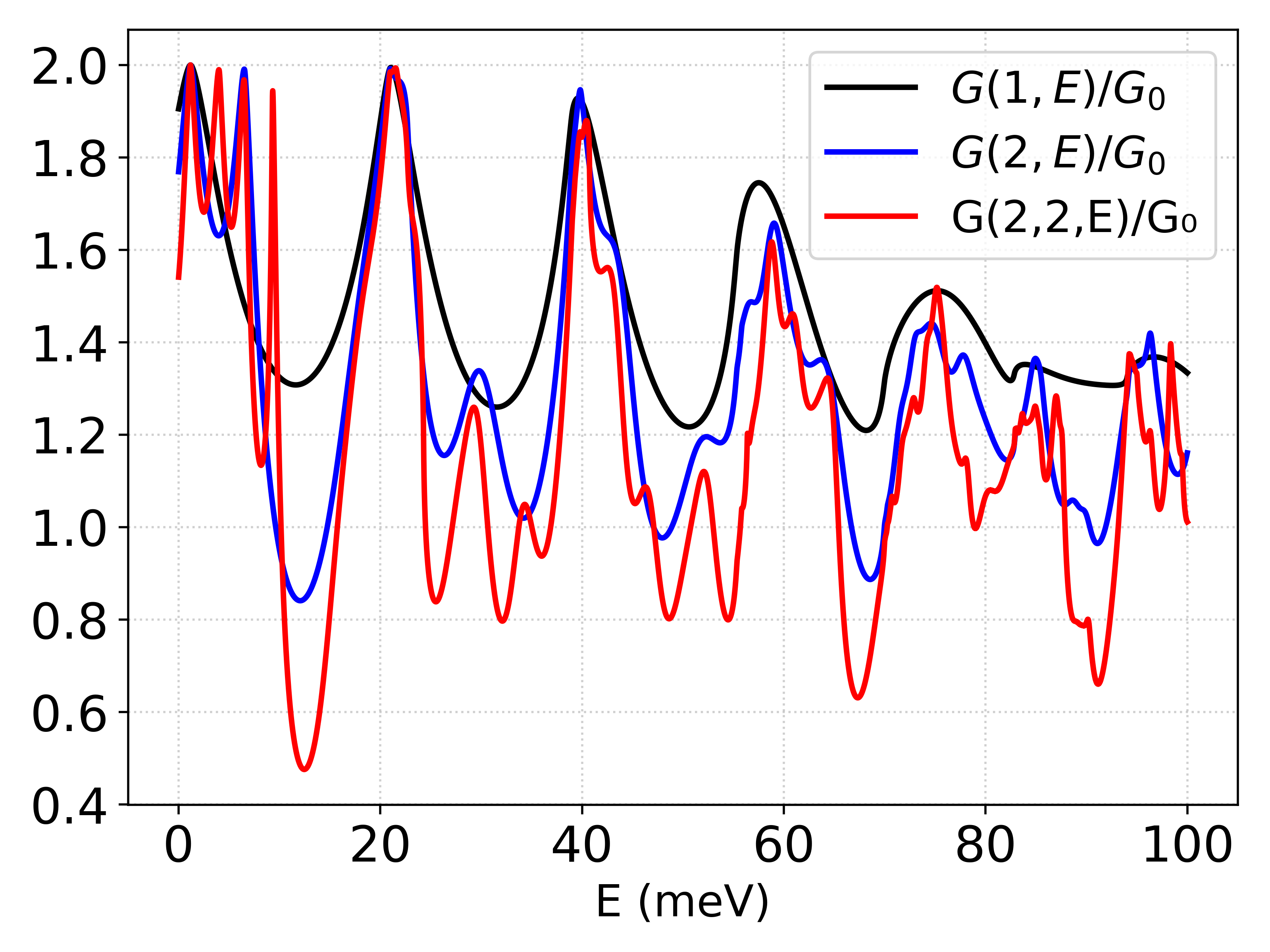}\label{fig:8b}}
\sg[~]{\ig[height=4cm]{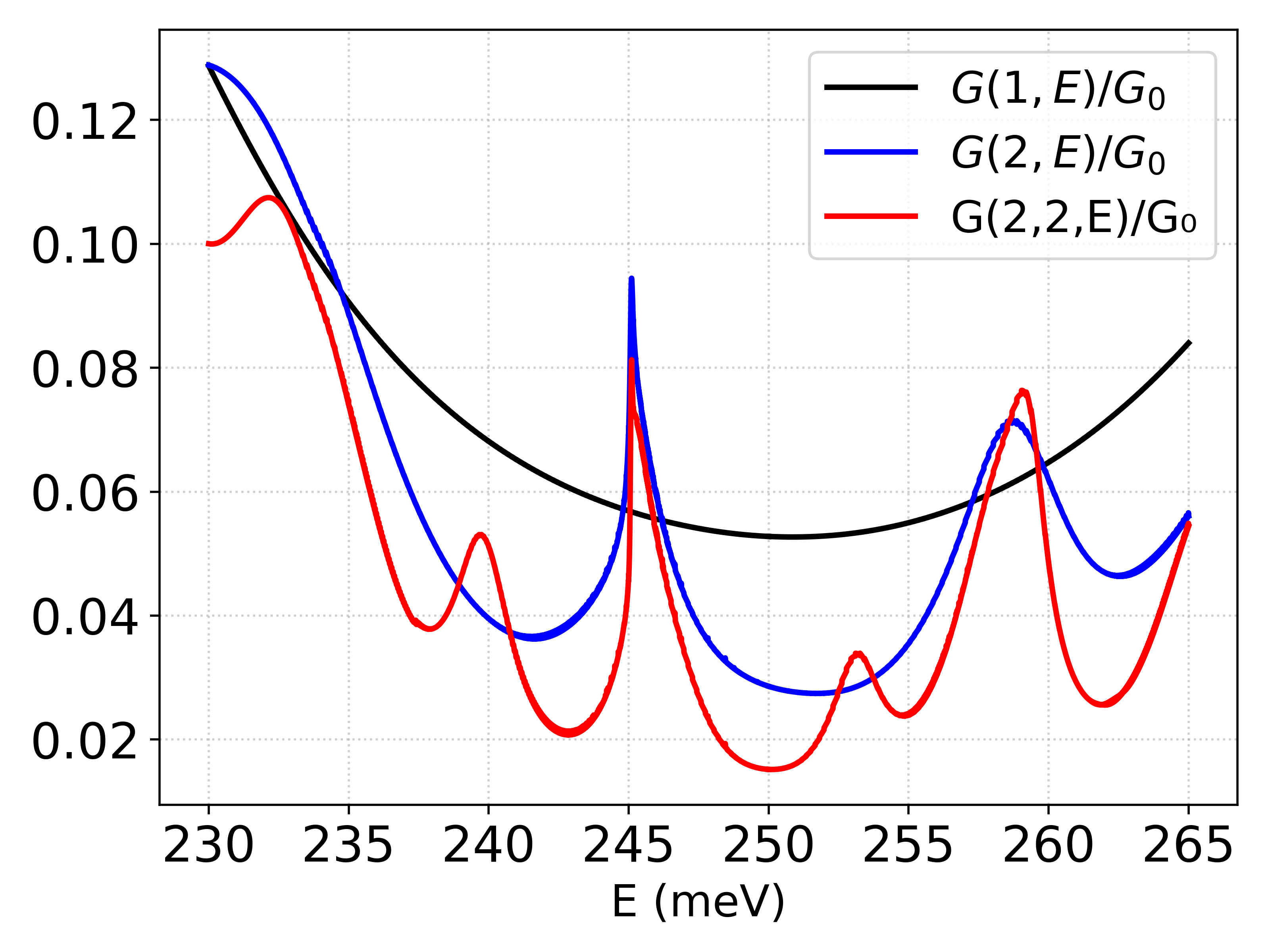}\label{fig:8c}}
\end{center}
\caption[conductance]{(a) The conductance, plotted against Fermi energy ($E$), is illustrated for three different cases: a single barrier (shown in black), a periodic potentials with $N_{1} = 2$ (shown in blue), and a SPPs of order $2$ with $N_{1} = 2$ and $N_{2} = 2$ (shown in red). All other parameters are the same as in FIG. \ref{fig:6}. Plots (b) and (c) are the sectional plot of (a). Plot (b) is evident that the period remains consistent across all curves. However, as the order increases, the amplitude of oscillations also increases. Near the Dirac point, the conductance tends to zero with the increasing order of periodicity, as evidenced by the plot (c).}
\label{fig:8} 
\end{figure}
\end{widetext}
\subsubsection{Conductance and shot noise:} In this subsection we explore the conductance in presence of periodic potentials and SPPs. The zero temperature Landauer-Buttiker formalism for conductances and the fano factors for the finite potential has been demonstrated in Ref. \cite{titov2007impurity}. In this section, we calculate the conductance and fano factor for periodic potentials and SPPs of order-$2$ having the expression of transmission probability in equations (\ref{periodic transmission}) and (\ref{spp transmission}). The Landauer-Buttiker conductances and the expressions for the fano factor are given for these cases as follows:
\begin{figure}[htb]
\begin{center}
\sg[~]
{\ig[height=3cm]{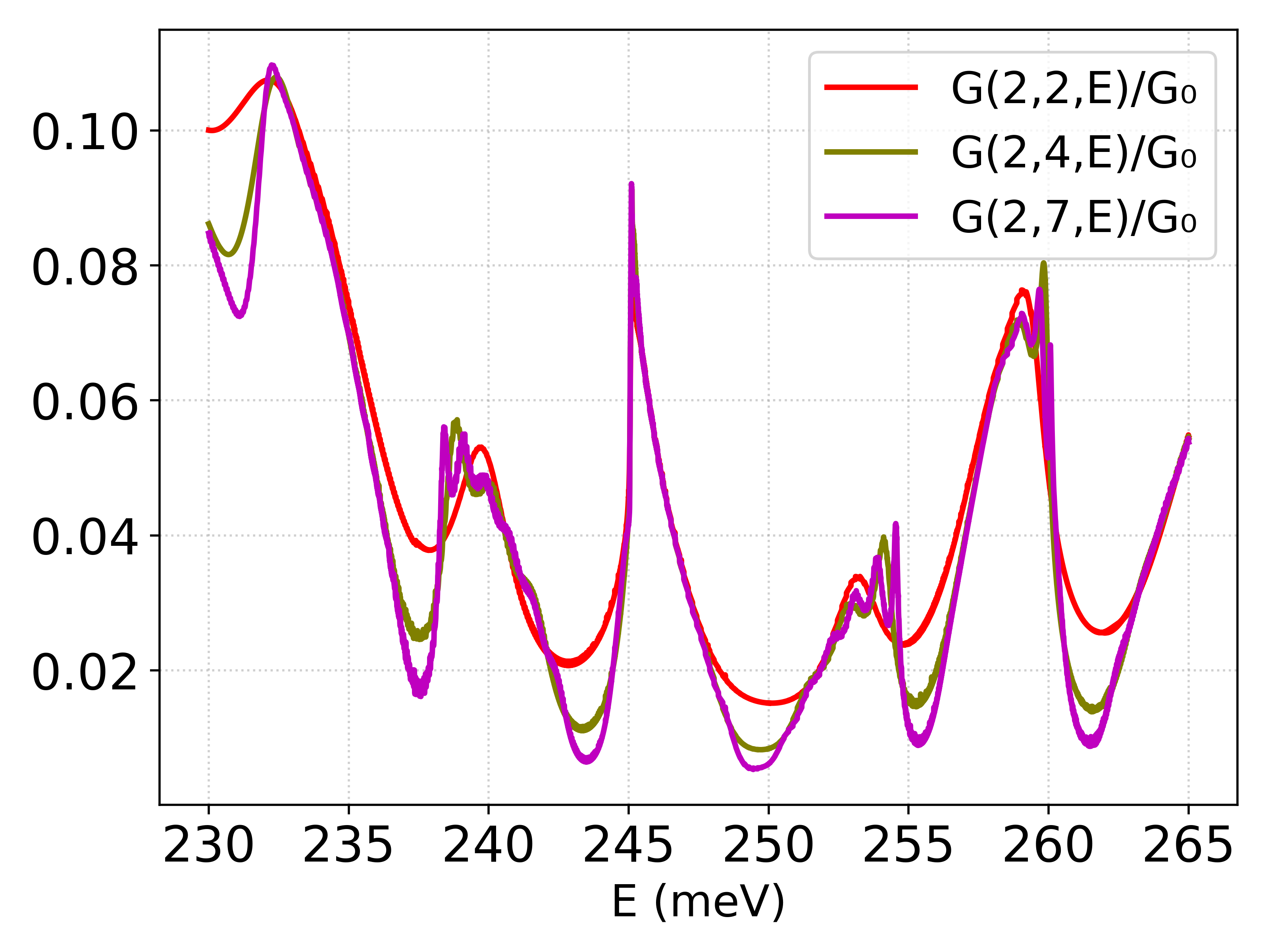}\label{fig:9a}}
\sg[~]{\ig[height=3cm]{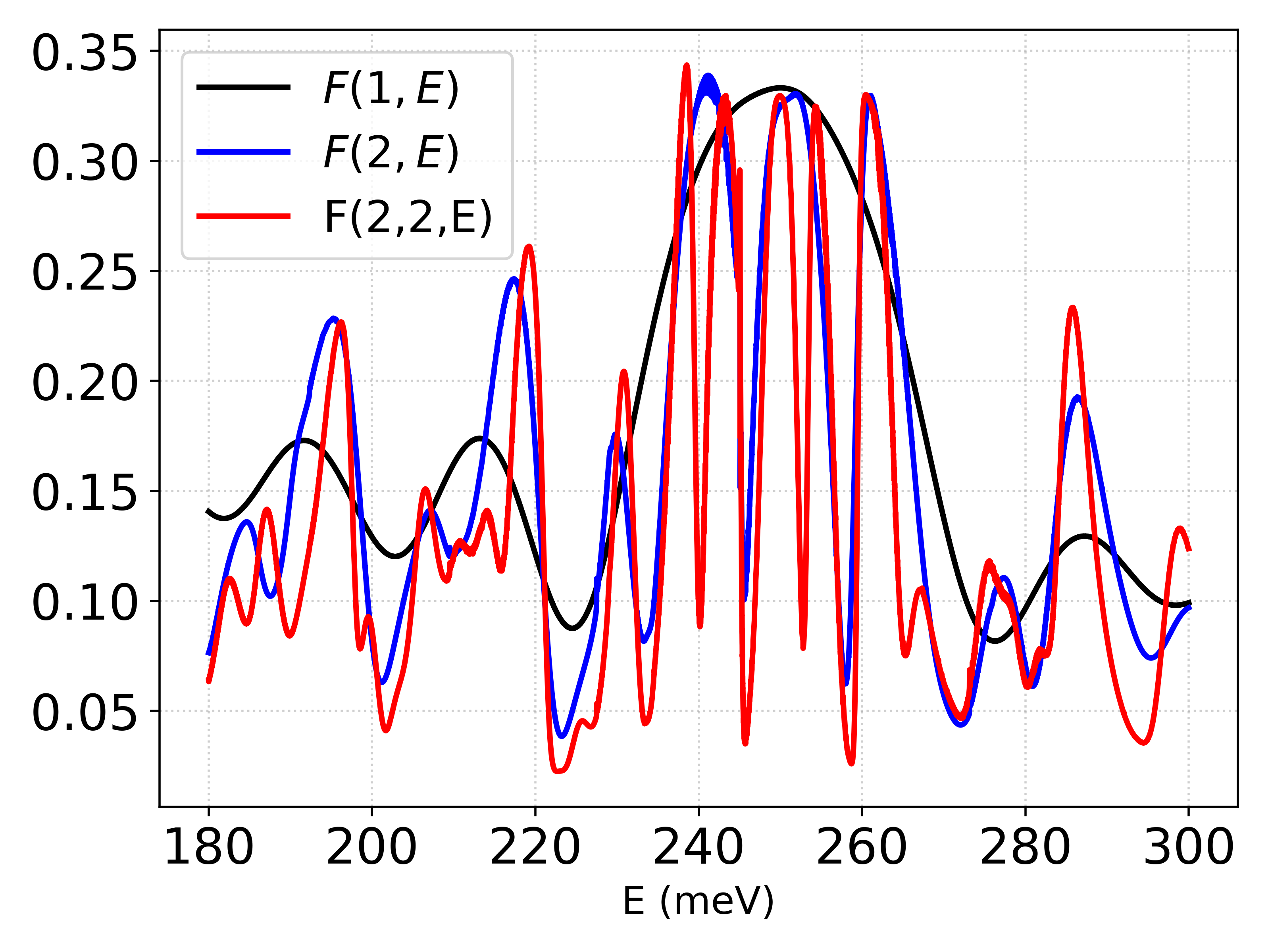}\label{fig:9b}}
\end{center}
\caption[Ribbon Geometry]{(a) The conductance, plotted against Fermi energy ($E$), is illustrated for three different $N_{2}$ values $N_{2}=2$, $N_{2}=4$, and $N_{2}=7$. (b) The Fano factor, plotted against Fermi energy ($E$), is illustrated for three different cases: the unit barrier (shown in black), a periodic potentials with $N_{1} = 2$ (shown in blue), and a SPPs of order $2$ with $N_{1} = 2$ and $N_{2} = 2$ (shown in red). All other parameters are the same as in FIG. \ref{fig:6}.} 
\label{fig:9} 
\end{figure}
\begin{eqnarray}
G(N_{1},E)&=&G_{0}(E)\int_{-\frac{\pi}{2}}^{\frac{\pi}{2}}{T_{1}\cos{(\phi)}}d\phi\\
G(N_{1},N_{2},E)&=&G_{0}(E)\int_{-\frac{\pi}{2}}^{\frac{\pi}{2}}{T_{2}\cos{(\phi)}}d\phi\\
F(N_{1},E)&=&\frac{\int_{-\frac{\pi}{2}}^{\frac{\pi}{2}}T_{1}(1-T_{1})\cos{(\phi)}d\phi}{\int_{-\frac{\pi}{2}}^{\frac{\pi}{2}}T_{1}\cos{(\phi)}d\phi}\\
F(N_{1},N_{2},E)&=&\frac{\int_{-\frac{\pi}{2}}^{\frac{\pi}{2}}T_{2}(1-T_{2})\cos{(\phi)}d\phi}{\int_{-\frac{\pi}{2}}^{\frac{\pi}{2}}T_{2}\cos{(\phi)}d\phi}
\end{eqnarray}
where $G_{0}(E)=\frac{2e^{2}EL_{y}}{\pi h}$ and $L_{y}$ represent the length of the graphene sample along the $y$-direction, and $T_{1}$, $T_{2}$ are the transmission probabilities for the periodic and SPPs of order-2 respectively.
The overall behavior of conductance for periodic and SPPs is similar to the single potential as shown in FIG. \ref{fig:8a}. However, in the case of periodic and SPPs, the conductance has an oscillatory behavior in each period of the single potential's conductance. As the order of periodicity increases, the oscillations become more pronounced as shown in FIG. \ref{fig:8b}. Also, the minimum non-zero conductance near the Dirac point $(E=V_{0})$ approaches zero increasing the order of periodicity. In this context, the super-periodic structure exhibits significantly lower conductivity near the Dirac point compared to both the periodic and unit barrier structures. This is demonstrated in FIG. \ref{fig:8c}. The conductance for super-periodic structure also approaches zero near the Dirac point as the number of periodicities $(N_{2})$ increases as shown in FIG. \ref{fig:9a}. In the case of a single barrier, at the Dirac point, the conductance assumes its minimal value, while the Fano factor reaches its peak of $\frac{1}{3}$. This observation aligns with the findings reported by Tworzydlo et al \cite{katsnelson2006zitterbewegung}. Furthermore, for both the periodic and super-periodic gap structures, the Fano factor converges towards $\frac{1}{3}$ at the Dirac point, accompanied by significant fluctuations in amplitude, as demonstrated in FIG. \ref{fig:9b}.
\section{Cantor Potentials}
\label{sec:IV}
In this section we extend our formalism to fractal potentials. One of the simplest examples of fractals is the Cantor set. This section demonstrates how some families of Cantor set potentials \cite{singh2023quantum2} can be thought of as super-periodic rectangular potentials in particular. Three Cantor set potentials types are covered in this article: UCPs, General cantor potentials, and GSVC set potentials. By employing the concept of super-periodicity, closed-form expressions for the transmission amplitude for SPPs are derived.
\subsection{UCPs}
\begin{figure}[htb]
\centering
\includegraphics[width=8cm]{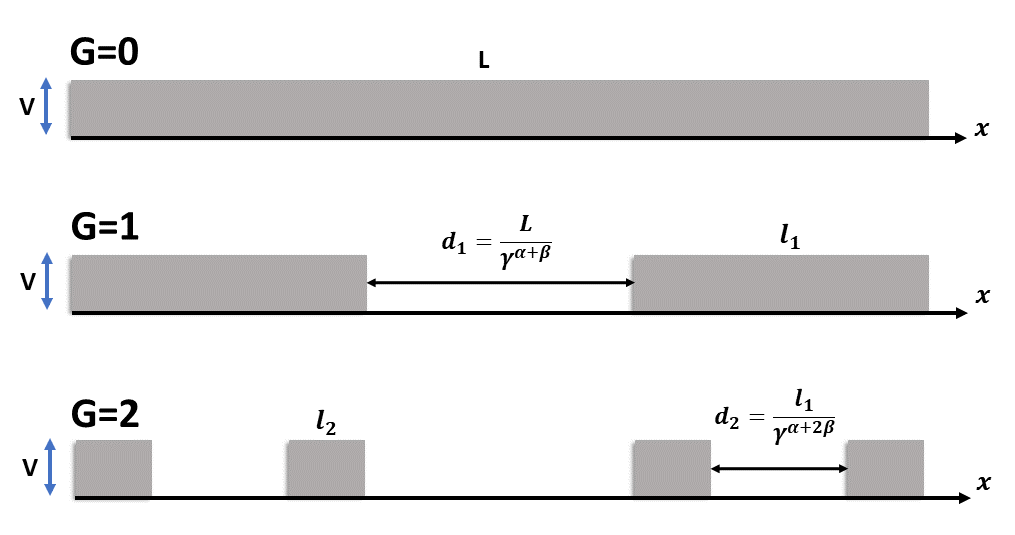}
\caption{The UCPs-$\gamma$ system is constructed with a visual representation wherein the white region shows the gap between potentials, and the height of the opaque segment corresponds to the potential height $(V)$. In this figure, $G$ denotes the system stage, while $l_{G}$ and $d_{G}$ represent the lengths of individual potential segments and the gap between potentials at the $G$th stage, respectively.}
\label{fig:10}
\end{figure}

In this subsection, we provide a brief overview of the main idea of the UCPs \cite{umar2023quantum, umar2024polyadic} for the sake of completeness. The UCPs-$\gamma$ system is constructed by starting with a rectangular barrier potential of finite length $L$ and height $V$, and iteratively removing a fraction $\frac{1}{\gamma^{\alpha + \beta G}}$ from the middle of each segment at each stage $G$. Here, $\gamma$ is a real positive number that is greater than $1$. $\alpha$ and $\beta$ are non-zero real numbers and they are not both zero at the same time. The structure of UCPs-$\gamma$ is demonstrated in FIG. \ref{fig:10}. In this figure, stage $G = 1$ corresponds to the removal of a fraction $\frac{1}{\gamma^{\alpha + \beta G}}$ from the center of a potential with height $V$ and length $L$. There are two possible segments of height $V$ and length $l_{1}$ in stage $G=1$. A fraction $\frac{1}{\gamma^{\alpha +\beta G}}$ is removed from the middle of the final two possible segments of stage $G=1$ for the subsequent stage, $G=2$, and in this instance, each potential segment has length $l_{2}$. For stage $G=3$, a fraction $\frac{1}{\gamma^{\alpha +\beta G}}$ is eliminated from the middle of the remaining four potential segments of stage $G=2$, and in this case, each potential segment has length $l_{3}$. Generally speaking, the UCPs-$\gamma$ system of any arbitrary stage $G$ consisting of $2^{G}$ potential segments of equal length $l_{G}$ is constructed by applying the same process of elimination of the fraction $\frac{1}{\gamma^{\alpha +\beta G}}$ from the middle of the remaining each potential segments of stage $G-1$.
It is observed from the UCPs-$\gamma$ system in FIG. \ref{fig:10}
\begin{eqnarray}
l_{G}&=&\frac{L}{2^{G}}\prod_{j=1}^{G}\left(1-\frac{1}{\gamma^{\alpha+\beta j}}\right)\label{lg};~~~~d_{G}=\frac{l_{G-1}}{\gamma^{\alpha+G\beta}}
\end{eqnarray}
and the super-periodic distance \cite{umar2023quantum},
\begin{equation}
s_{p}=\frac{L}{2^{G+1-p}}\left(1+\frac{1}{\gamma^{\alpha+\beta(G+1+p)}}\right)\prod_{j=1}^{G-p}\left(1-\frac{1}{\gamma^{\alpha+\beta j}}\right)
\end{equation}
with $p=1,2,\ldots,G$.
UCPs-$\gamma$ are a special case of super-periodic rectangular potentials and the transfer matrix of ‘unit cell’ rectangular barrier for relativistic particle of width $l_{G}$ and height $V$ can be written using equations (\ref{m11kg}) and (\ref{m12kg}), as
\begin{eqnarray}
m_{11}&=&(\cos(q^{\prime} l_{G})-i \epsilon_{+} \sin(q^{\prime} l_{G}))e^{iq l_{G}}=m_{22}^{*}\\
m_{12}&=&i \epsilon_{-} \sin(q^{\prime} l_{G})=m_{21}^{*}
\end{eqnarray}
Therefore, the final expression for the transmission coefficient of the UCPs-$\gamma$ system is obtained as:
\begin{equation}
T_{G}=\frac{1}{1+4^{G}\epsilon_{-} \sin(q^{\prime} l_{G})\prod_{j=1}^{G}\xi_{j}^{2}}\label{UCP}
\end{equation}
where,
\begin{eqnarray}
\xi_{j}&=&2^{j-1}|m_{22}|\cos(\theta-q \rho_{1})\prod_{p=1}^{j-1}\xi_{p}\nonumber\\
&&-\sum_{r=1}^{j-1}\left(2^{j-r-1} \cos(q \rho_{2})\prod_{p=r+1}^{j-1}\xi_{p}\right)\label{bbloch}\\
\theta&=&\tan^{-1}(\xi_{+}\tan(q^{\prime}l_{G}))-q l_{G}
\end{eqnarray}
\begin{eqnarray}
\rho_{1}&=&\left(\sum_{p=1}^{j-1}s_{p}\right)-s_{j}\\
\rho_{2}&=&\left(\sum_{p=r}^{j}s_{j}\right)-(2 s_{j}-s_{r})
\end{eqnarray}
In the non-relativistic limit our results match with the results in the literature \cite{hasan2018super}.
And as $V\to\infty$, the transmission probability has the form:
\begin{equation}
T_{G}\to\frac{1}{1+4^{G}V \sin(V l_{G})\prod_{j=1}^{G}\Xi_{j}^{2}}
\end{equation}
where $\Xi$ is the argument of the CP at $V\to\infty$. This indicates that when $V l_{G}=\omega\pi$, with $\omega$ being an integer, or when the value of CP is zero, the transmission coefficient equals $1$ for any order. Below, we discuss the properties of the transmission coefficient for a few special cases.
\subsubsection{Case 1: General cantor potentials $(\alpha=1,~\beta=0)$}
 General Cantor fractal system is a particular instance of UCPs-$\gamma$ system. Thus, for a general Cantor fractal system, length $l_{G}$ of the potential segment at any given stage $G$ can be expressed using Eq.(\ref{lg}).
\begin{equation}
l_{G}=L\left(\frac{\gamma-1}{2\gamma}\right)^{G}
\end{equation}
The equation (\ref{bbloch}) can be used to define the Bloch phase for the General Cantor fractal system, and Equation (\ref{UCP}) can be used to express the tunneling probability based on the previously specified parameters.
\onecolumngrid
\begin{widetext}
\begin{figure}[htb]
\begin{center}
\sg[~]{\ig[height=4.2cm]{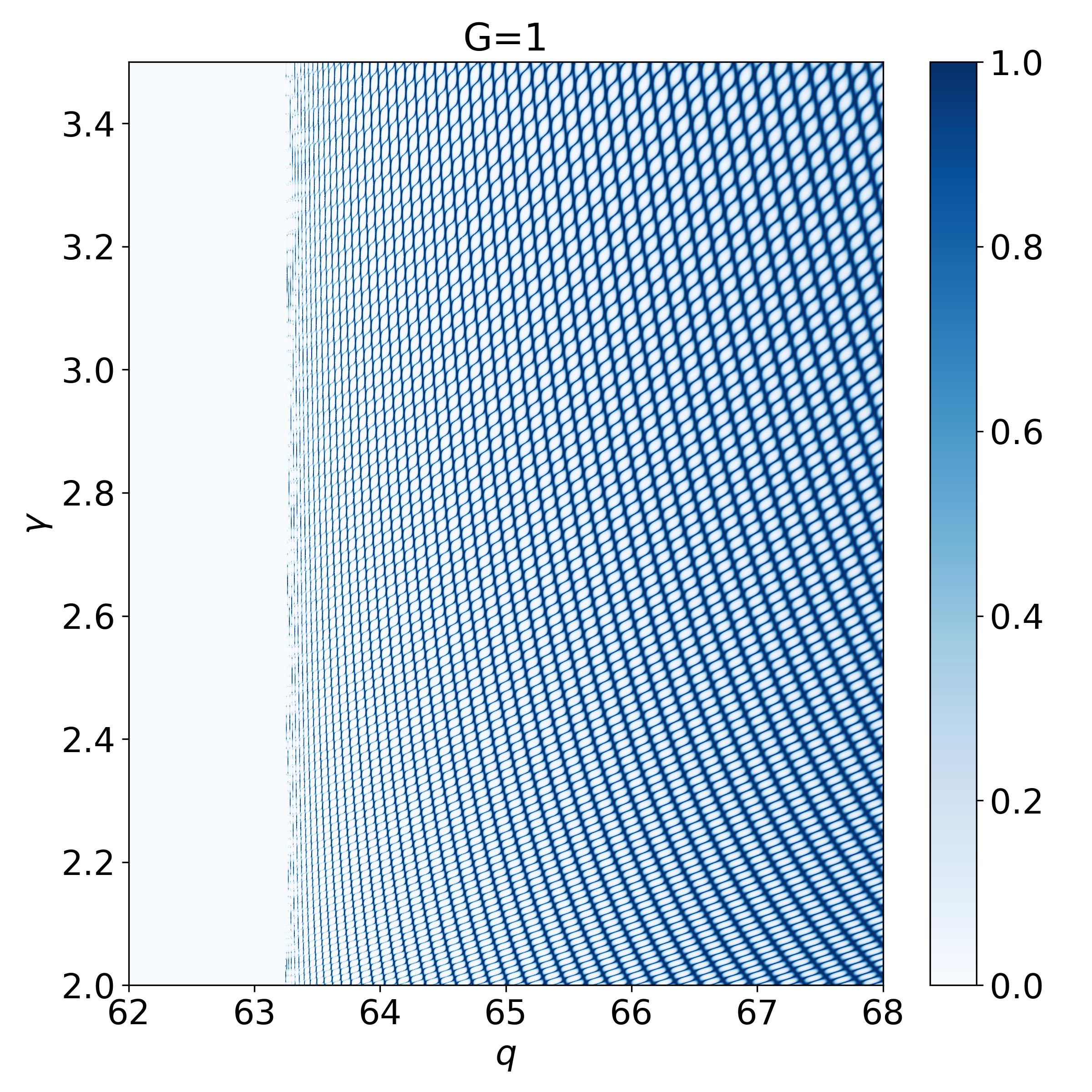}\label{fig:11a}}
\sg[~]{\ig[height=4.2cm]{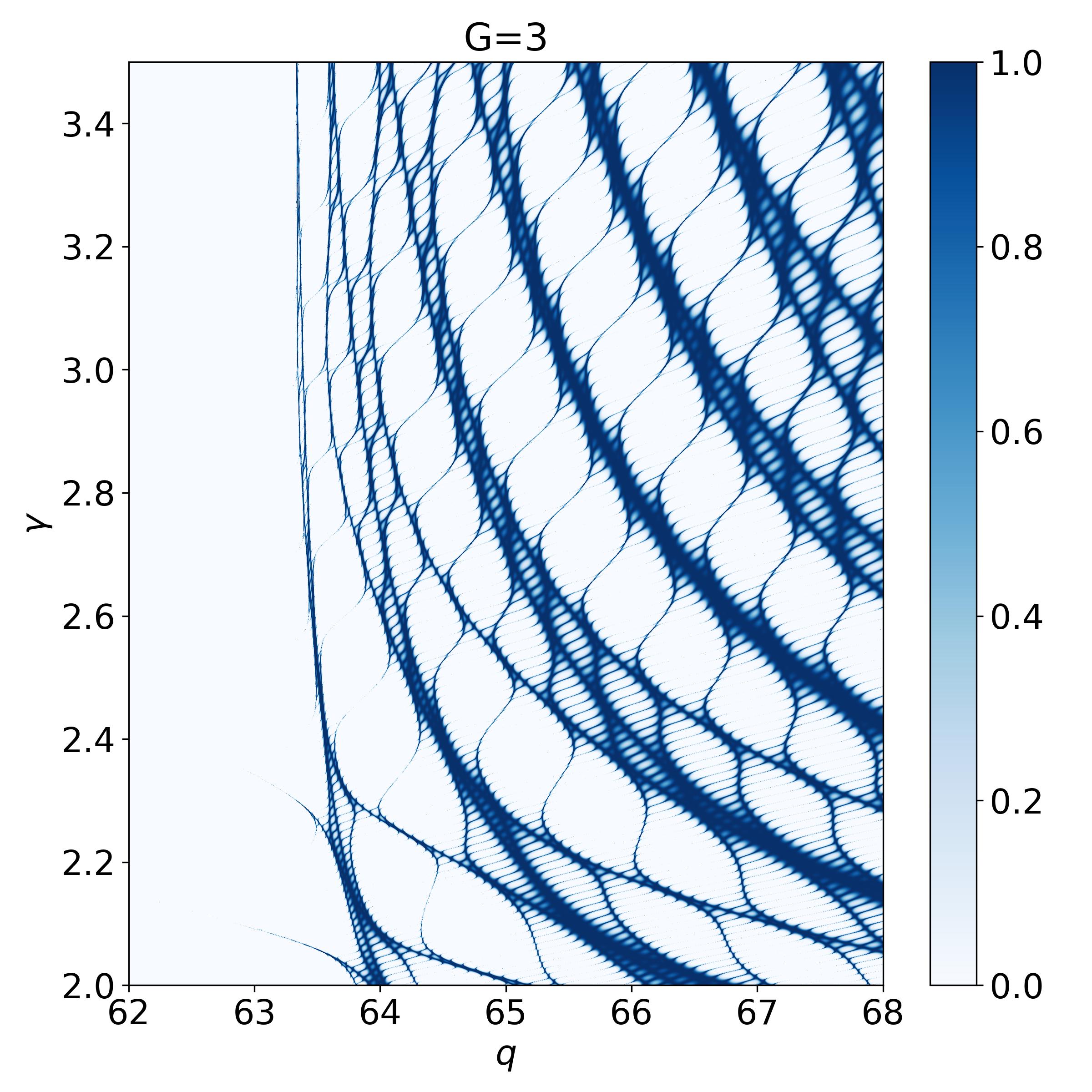}\label{fig:11b}}
\sg[~]{\ig[height=4.2cm]{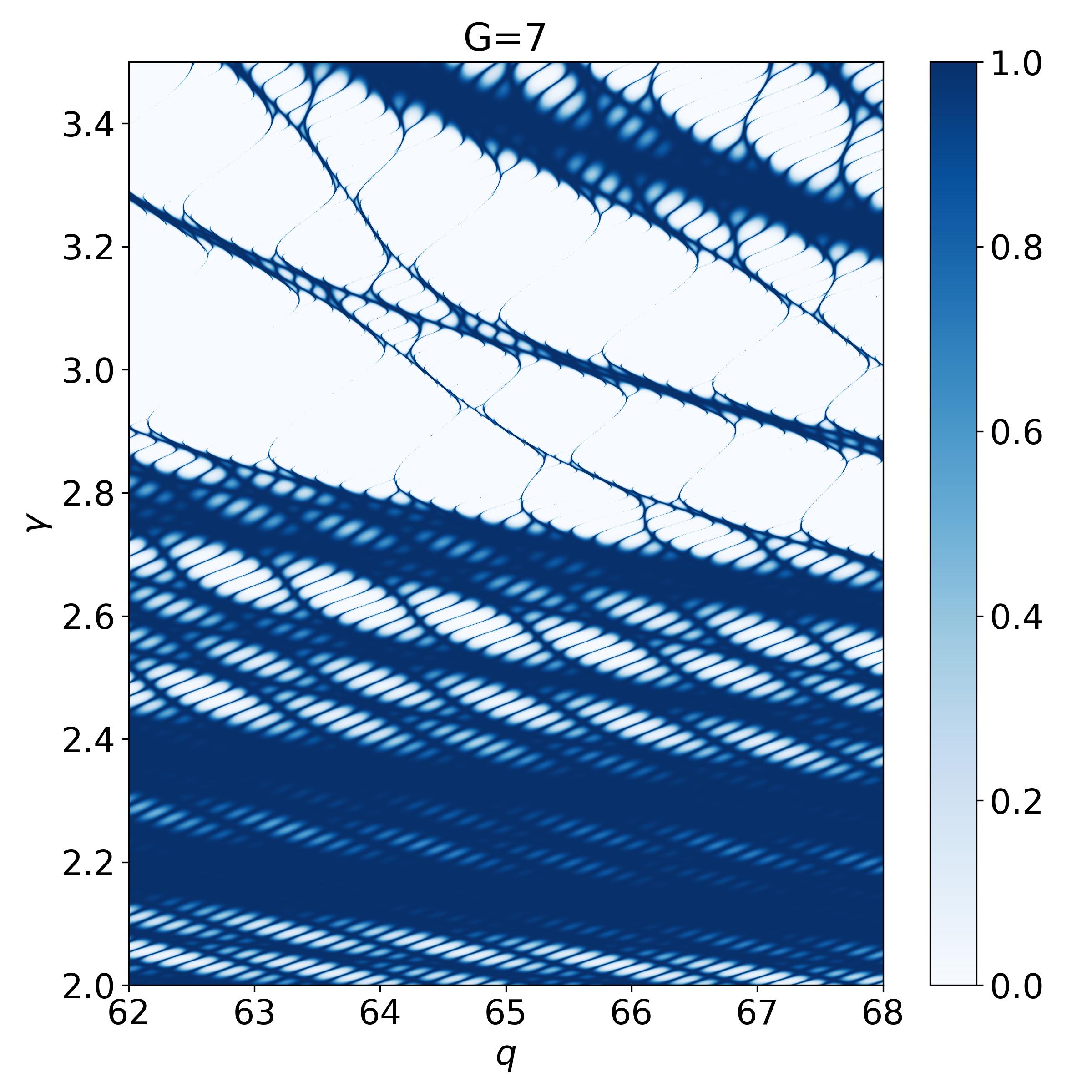}\label{fig:11c}}
\sg[~]{\ig[height=4.2cm]{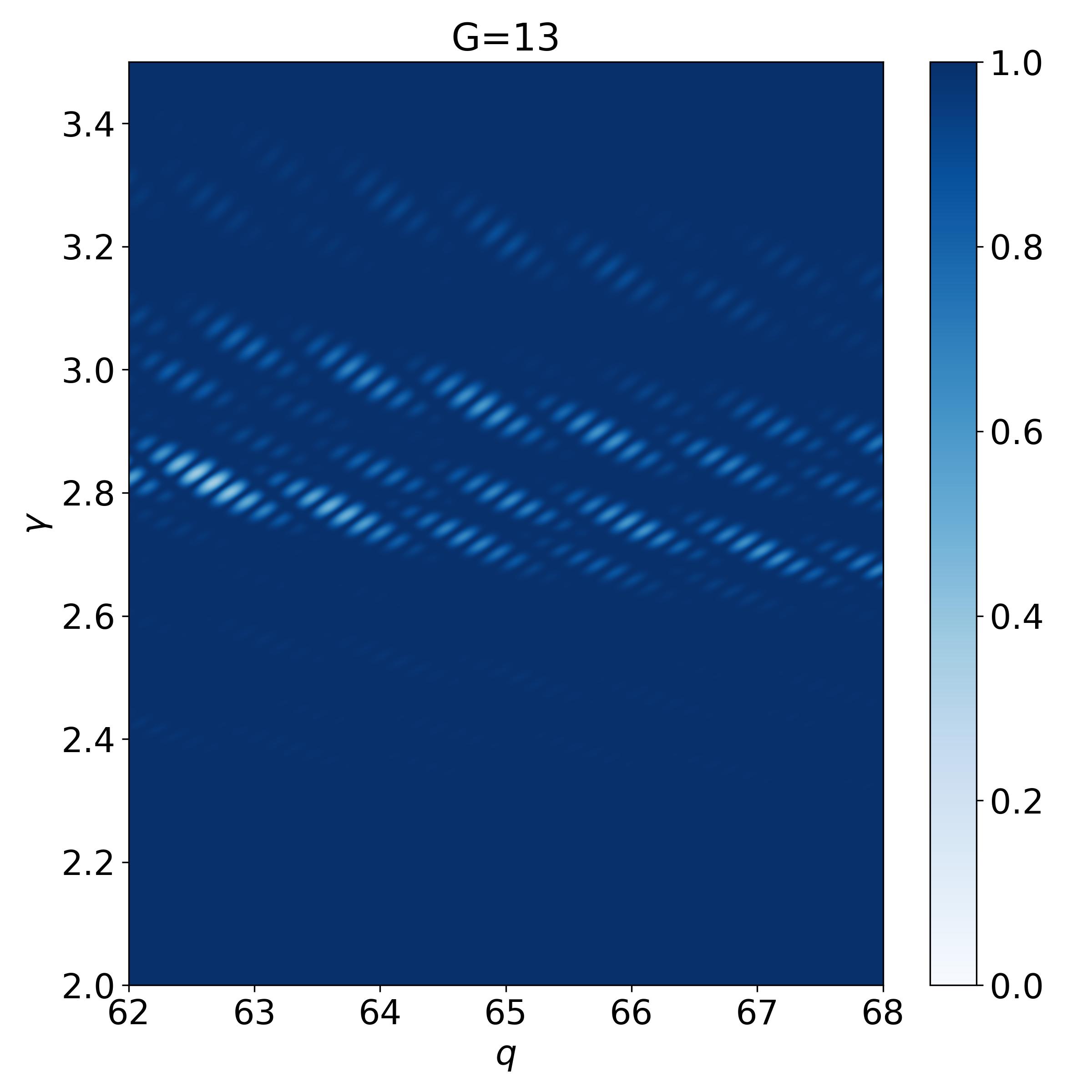}\label{fig:11d}}
\end{center}
\caption[Bulk, edge and corner state energies]{Density plot of tunneling coefficient $T$ for different stages $G$ of the General Cantor Potentials of height $V = 2000$, $m=1$ and total span $L = 20$.}
\label{fig:11} 
\end{figure}
\end{widetext}
FIG. \ref{fig:11} displays the density plot of transmission probability for a general Cantor potentials for relativistic particles of multiple stages $G$ with changing $\gamma$ and $q$. These plots display very sharp peaks of the tramsmission at special values of $\gamma$ and $q$, corresponding to the condition $Vl_{G} = \omega \pi$ and the zeros of the CPs. The unit cell rectangular potential gets thinner as $G$ increases, making it easier for the wave to transmits the fractal system. For $\gamma=3$, whole system behaves as standard cantor potentials system.
\subsubsection{Case 2: GSVC potentials $(\alpha=0,~\beta=1)$}
GSVC is also an special case of UCPs-$\gamma$ system \cite{singh2023quantum, narayan2023tunneling}. When $\alpha$ and $\beta$ are substituted in Eq. (\ref{lg}), $l_{G}$ for general SVC is expressed as follows:
\begin{equation}
l_{G}=\frac{L}{2^{G}}\prod_{j=1}^{G}\left(1-\frac{1}{\gamma^{j}}\right)
\end{equation}
Similarly, the Bloch phase for the GSVC fractal system can be defined using equation (\ref{bbloch}), and the tunneling probability based on the previously given parameters can be expressed using equation (\ref{UCP}).
Also, For $\gamma=4$, whole system behaves as standard SVC potentials system.
\onecolumngrid
\begin{widetext}
\begin{figure}[htb]
\begin{center}
\sg[~]{\ig[height=4.3cm]{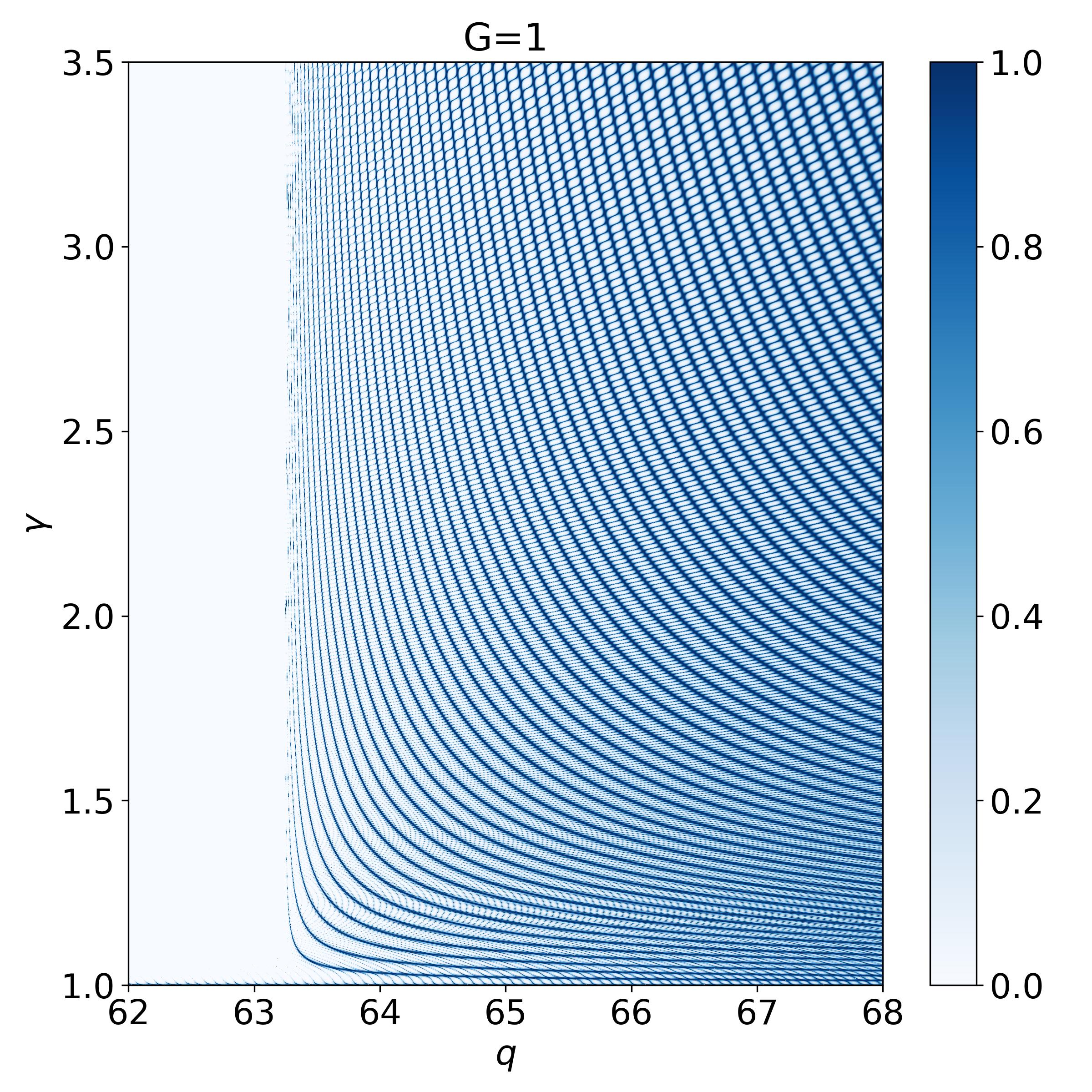}\label{fig:12a}}
\sg[~]{\ig[height=4.3cm]{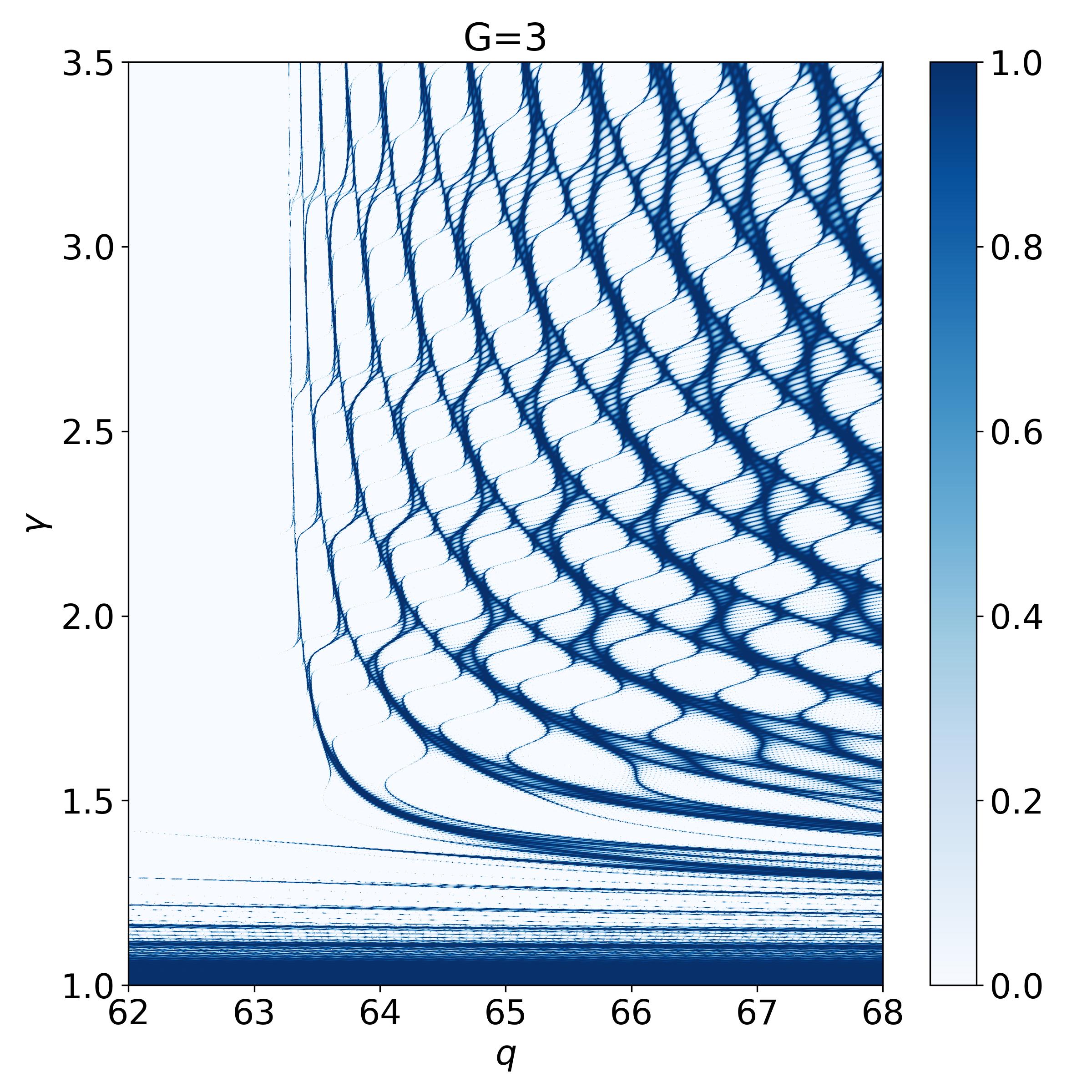}\label{fig:12b}}
\sg[~]{\ig[height=4.3cm]{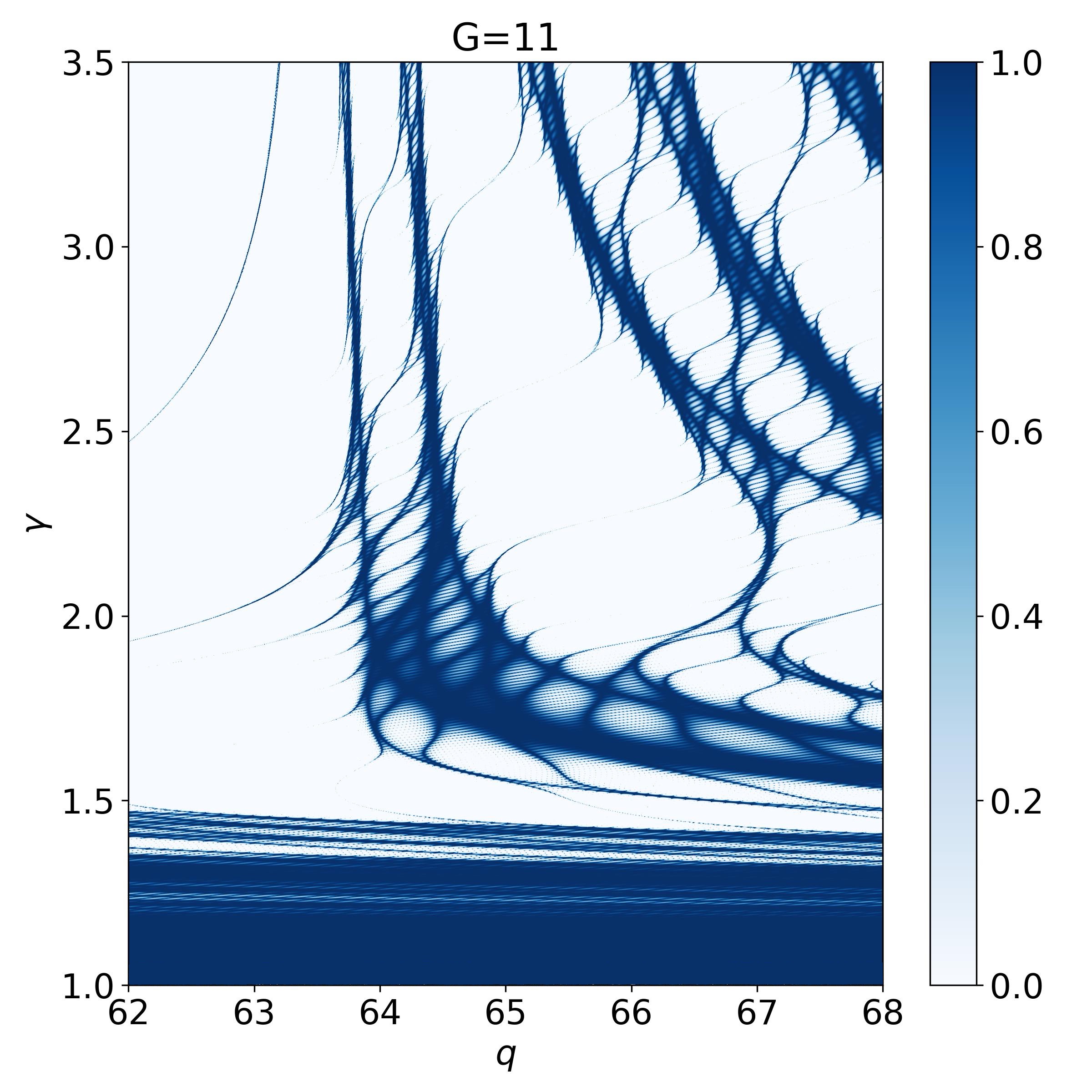}\label{fig:12c}}
\sg[~]{\ig[height=4.3cm]{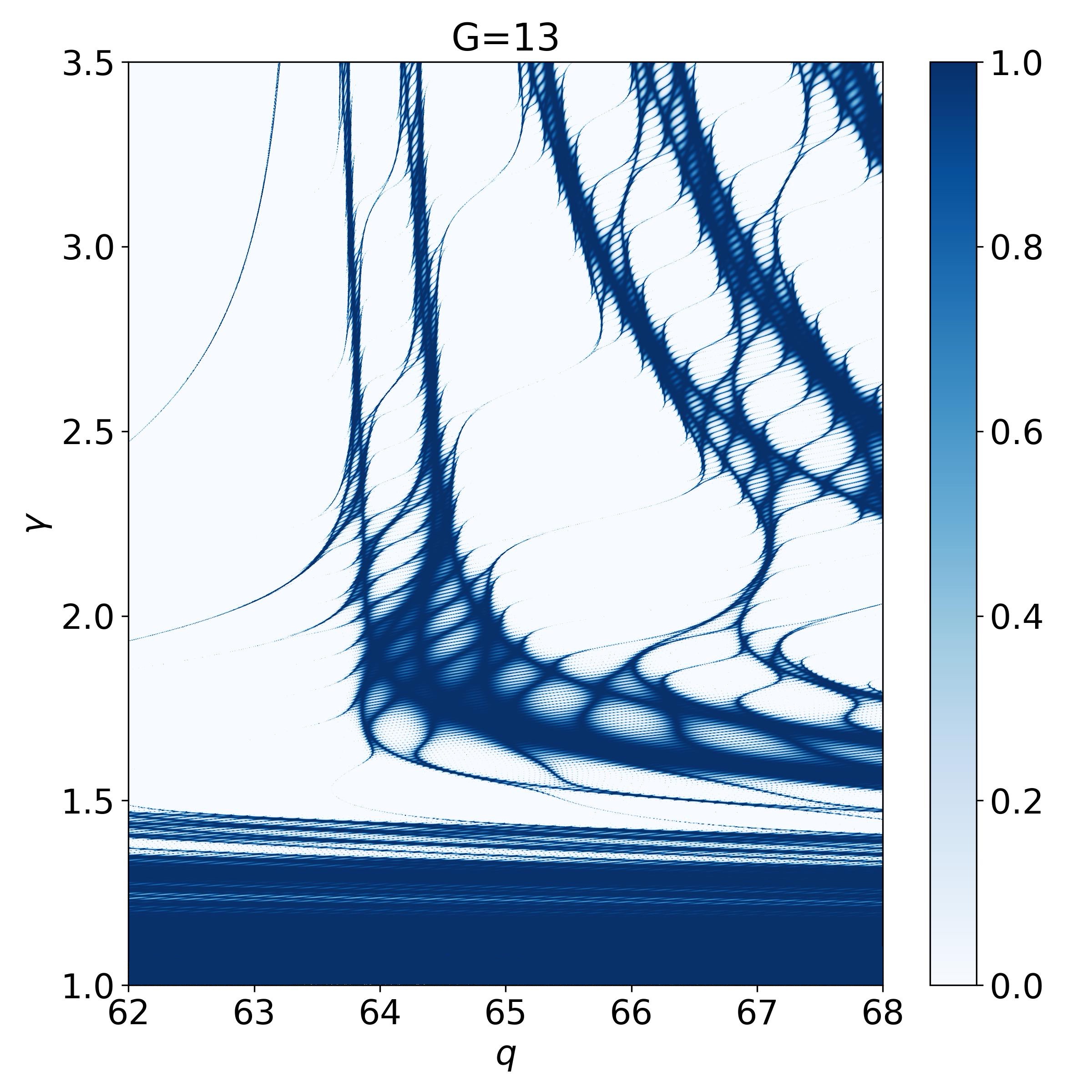}\label{fig:12d}}
\end{center}
\caption[Bulk, edge and corner state energies]{Density plot of tunneling coefficient $T$ for different stages $G$ of the GSVC potentials of height $V = 2000$, $m=1$ and total span $L = 20$.}
\label{fig:12} 
\end{figure}
\end{widetext}
The density plot of tunneling coefficient $T_{G}$ for different stages $G$ of the GSVC potentials is shown in FIG. \ref{fig:12} . It can be shown from this figure that the tunneling coefficients are almost equal to unity for all wave numbers $q$ near $\gamma\approx 1$. This is because a greater percentage of the entire span $L$ gets eliminated when $\gamma$ approaches unity. There is absolutely no barrier when $\gamma=1$. Furthermore, at a slightly larger value of $\gamma$ beyond $1$, we found more transparency, since more fraction of the GSVC potentials is removed as $G$ increases. It is also observed in FIG. \ref{fig:12}  that the plots for transmission coefficients for $G = 11$ and $G = 13$ are almost identical. This is because, at each stage $G$, a progressively less portion of the earlier segments are eliminated; as a result, the transmission coefficient would exhibit a saturation with rising GSVC potentials stage. When $\gamma$ values are higher, this will become more apparent.
\section{Conclusion}\label{sec:V}
In this article, we have used the transfer matrix method to study the tunneling behavior of relativistic particles encountering SPPs. We have focused on following key aspects: the general behavior of spinless Klein particles and the specific case of massless Dirac electrons in monolayer graphene. We have analytically derived the reflection and transmission probabilities when they encounter rectangular potential barriers arranged in a super-periodic pattern. We have found that these relativistic particles exhibit a significantly higher degree of reflection compared to their non-relativistic counterparts in the certain energy range. Interestingly, we have also demonstrated the existence of finite transmission probability, even for infinitely large super-periodic rectangular SPPs; a unique characteristic of the Klein tunneling.
Focusing on monolayer graphene, we have studied how massless Dirac electrons behave when they encounter super-periodic rectangular potential barriers. We have thoroughly analyzed the transmission probability, conductance, and Fano factor based on important factors such as the number of barriers, the order of super-periodicity, and the angle of incidence. The transmission probability has shown a series of resonances that are closely related to the number of barriers and the super-periodicity order.\\
The number of electrostatic barriers has significantly impacted the transmission probability for both periodic and SPPs. For normal incidence ($\phi = 0^\circ$), the transmission coefficient has remained unity, independent of the number of electrostatic barriers, confirming the Klein-tunneling effect. This has indicated complete transparency of the system at normal incidence, irrespective of barrier width. As the incidence angle $\phi$ has increased, the transmission coefficient peaks have become sharper. With an increase in the number of barriers ($N_{1} > 1$), the transmission probability has developed additional groups of peaks, where each group has contained $N_{1} - 1$ peaks within specific incidence angle ranges. For the super-periodic electrostatic potentials of order-2, the transmission behavior has been analogous to the periodic potentials but has featured $N_{2}$ resonance peaks within each periodic peak, making it challenging to distinguish individual resonance peaks for higher $N_{2}$.\\
These $N_{2}$ resonance peaks in the SPPs of second order have been explained by the CP $U_{N_{2}-1}(\xi_{2})$ governing the transmission coefficient. Since a CP of degree $N_{2}-1$ has had $N_{2} - 1$ roots within the interval from $-1$ to $1$, and an additional peak has arisen from $U_{N_{1}-1}(\xi_{1}) = 0$, the total number of resonance peaks has become $N_{2}$.\\

Further, the conductance and shot noise for periodic potentials and SPPs have been analyzed using the Landauer-Buttiker formalism. The expressions for conductance and Fano factors have highlighted the oscillatory nature of conductance within each period of the single potential's conductance. As the periodicity has increased, these oscillations have become more pronounced, with the minimum non-zero conductance near the Dirac point ($E = V_{0}$) approaching zero. Notably, the super-periodic structures have exhibited significantly lower conductivity near the Dirac point compared to both periodic and single-barrier structures. This trend has been evident in the conductance behavior as the number of periodicities $(N_{2})$ has increased. At the Dirac point, the conductance for a single barrier has reached its minimal value while the Fano factor has peaked at $\frac{1}{3}$, consistent with previous findings by Tworzydlo et al. \cite{katsnelson2006zitterbewegung}. For both periodic and super-periodic gap structures, the Fano factor has also converged to $\frac{1}{3}$ at the Dirac point, with substantial amplitude fluctuations. These results have provided a comprehensive understanding of how periodicity and super-periodicity have affected the electronic transport properties in graphene-based systems.\\
Extending our formalism to fractal potentials, such as Cantor set potentials, has further demonstrated the versatility and applicability of the super-periodic concept. UCPs, General Cantor potentials, and GSVC potentials have been explored, revealing that these fractal structures can be treated as super-periodic rectangular potentials. The closed-form expressions for the transmission amplitude of these potentials have highlighted how fractal geometry has influenced the behavior of SPPs. For instance, the UCP-$\gamma$ system has exhibited $G$-stage transmission characteristics where the transmission coefficient has equaled unity under specific conditions, as demonstrated with CPs.\\
The Cantor and SVC potentials have exhibited sharp transmission peaks that have become more prominent with increasing stages, as depicted in FIG. \ref{fig:11} and FIG. \ref{fig:12}. Notably, for $\gamma \approx 1$, the GSVC potentials has shown near-unity tunneling coefficients, reflecting the reduced barrier presence.\\

In conclusion, we have examined the behavior of a relativistic particle in the presence of a SPPs. We have derived exact analytical expressions for the transmission and reflection coefficients when a relativistic particle has encountered a super-periodic rectangular barrier. Our observations have shown that the transmission coefficient has displayed both the Klein tunneling and transmission resonance. Furthermore, we have derived exact analytical expressions for the transmission and reflection coefficients of electrons tunneling through SPPs in graphene. Finally, we have derived the reflection and transmission coefficients when barrier potentials have been arranged in the form of a fractal potentials. Our work has opened up several directions for future research, such as experimentally realizing this type of potential to investigate reflection and transmission behaviors, as well as exploring potential device applications.
\section{Acknowledgement}
A.D. acknowledges financial support from the SERB-SRG grant SRG/2022/001145 and the IoE seed grant from BHU IoE/Seed Grant II/2021-22/39963. BPM acknowledges the incentive research grant for faculty under the IoE Scheme (IoE/Incentive/2021-22/32253) of Banaras Hindu University. S.S. acknowledges NBCFDC, New Delhi, India for JRF fellowship.
 
\vspace{.5cm}
\appendix
\section{General Theory} \label{sec:app}       
Here, we explain the transfer matrix method for studying particle scattering from a potential barrier, considering both one-dimensional and a special case of two-dimensional potential where in one direction we have finite potential barrier which is extended to infinity in other direction. $y$-component of momentum is conserved for two-dimensional potential, and hence calculation is simplified.
\subsection{For one-dimensional scattering}
In this section, we discuss the transfer matrix method for studying the one-dimensional scattering of quantum particles from a potential barrier, represented by a potential $V(x)$ for $-a < x < a$ as shown in FIG. \ref{fig:1}. The general form of the time-independent wave function can be expressed as:
\begin{equation}
\psi(x) = \begin{cases} 
Ae^{ikx}+Be^{-ikx} &\mbox{ for } x<-a\\
\phi(x) & \mbox{ for } -a<x<a\\
Ce^{ikx}+De^{-ikx} & \mbox{ for }  x>a 
\end{cases}
\label{wavefunction}
\end{equation}
%
%
%
where $k$ is the wave vector and $\phi(x)$ depends on the behavior of the potential.
Using boundary conditions of the wave functions in equation ($\ref{wavefunction}$), we obtain two linear relationships among the coefficients $A$, $B$, $C$, and $D$. These relationships can be solved to express any two amplitudes in terms of the other two, resulting in a matrix equation that fully characterizes the transmission behavior of the system. The system of equations in matrix form is
\begin{eqnarray}
\begin{pmatrix}A\\B\end{pmatrix}=M\begin{pmatrix}C\\D\end{pmatrix}
\end{eqnarray}
The matrix $M$ is a $2\times2$ matrix.
\begin{eqnarray}
M=\begin{pmatrix}m_{11}&m_{12}\\m_{21}&m_{22}\end{pmatrix}
\label{1}
\end{eqnarray}
This is known as the ``transfer matrix".
The time reversal invariance, along with the conservation of probability imposes a relation between the elements of the transfer matrix for all hermitian scalar potential $V(x)$,
\begin{eqnarray}
m_{11}=m_{22}^{*}; ~\quad m_{12}=m_{21}^{*}
\label{}
\end{eqnarray}
and,
\begin{eqnarray}
|m_{11}|^{2}=1+|m_{12}|^{2}
\label{}
\end{eqnarray}
As a result, the transfer matrix $M$ is shown to be unimodular. By utilizing these properties, we can express the transmission and reflection probabilities for the scalar potential $V(x)$ in the following manner (if there are no particles moving backward on the right-hand side of the potential, i.e. $D=0$),
\begin{equation}
T=\frac{1}{|m_{11}|^{2}}; ~\hspace{1cm} R=\frac{|m_{12}|^{2}}{|m_{11}|^{2}}
\end{equation}
\subsubsection{Generalizing the transfer matrix method to the case of locally periodic potentials}
We use the formalism established by Griffiths \cite{griffiths_periodic} for the scattering of waves by a locally periodic medium, for that we consider the periodic potentials illustrated in FIG. \ref{fig:2}, where the potential $V(x)$ of a single `unit cell' is repeated periodically $N_{1}$ times over a distance $s_{1}$ along the $x$ direction (with $s_{1}=d_{0}+c_{1}$, $d_{0}=2a$, and $c_{1}>0$.
If we know the transfer matrix for the unit cell potential, then we can obtain the transfer matrix for the entire periodic system (consisting of an array of $N_{1}$ unit cells) as
\begin{widetext}
\begin{equation}
M(N_{1})=
\begin{bmatrix}
(m_{11}e^{-iks_{1}}U_{N_{1}-2}(\xi_{1})-U_{N_{1}-2}(\xi_{1}))e^{ikN_{1}s_{1}}&m_{12}U_{N_{1}-1}(\xi_{1})e^{-ik(N_{1}-1)s_{1}}\\
m_{12}^{*}U_{N_{1}-1}(\xi_{1})e^{ik(N_{1}-1)s_{1}}&(m_{11}^{*}e^{iks_{1}}U_{N_{1}-2}(\xi_{1})-U_{N_{1}-2}(\xi_{1}))e^{-ikN_{1}s_{1}}
\end{bmatrix}
\end{equation}
\end{widetext}
where $k$ is the wave vector, and $U_{N_{1}}(\xi_{1})$ is the CP of the second kind with the argument $\xi_{1}$, and
\begin{eqnarray}
\xi_{1}&=&\frac{1}{2}(m_{11}e^{-iks_{1}}+m_{11}^{*}e^{iks_{1}})\nonumber\\
&=&\text{Re}(m_{11})\cos{(ks_{1})}+\text{Im}(m_{11})\sin{(ks_{1})}
\end{eqnarray}
%
With the help of the transfer matrix, the transmission probability and the reflection probability for a locally periodic system are defined as,
\begin{eqnarray}
T&=&\frac{1}{1+(|m_{12}|U_{N_{1}-1}(\xi_{1}))^{2}}\\
R&=&\frac{(|m_{12}|U_{N_{1}-1}(\xi_{1}))^{2}}{1+(|m_{12}|U_{N_{1}-1}(\xi_{1}))^{2}}
\label{13a}
\end{eqnarray}
\subsubsection{Extension to SPPs}
The closed-form expression for the transfer matrix of SPPs $V_{n}$, in terms of the transfer matrix of the unit cell potential $V(x)$, is given by:
\begin{equation}	M(N_{1},N_{2},....,N_{n})=\begin{pmatrix}(m_{11})_{n}&(m_{12})_{n}\\(m_{21})_{n}&(m_{22})_{n}\end{pmatrix}
\end{equation}
The elements of the matrix are,
\begin{eqnarray}
(m_{11})_{n}&=& m_{11}\prod_{r=1}^{n}(U_{N_{r}-1}(\xi_{r}))e^{ik\sum_{r=1}^{n}(N_{r}s_{r}-s_{r})}\nonumber\\
&&-\sum_{r=1}^{n-1}(X_{r})_{n}-U_{N_{n}-2}(\xi_{n})e^{ikN_{n}s_{n}}=(m_{22})_{n}^{*}\nonumber\\
\end{eqnarray}
\begin{eqnarray}
(m_{12})_{n}&=& m_{12}\prod_{r=1}^{n}(U_{N_{r}-1}(\xi_{r}))e^{-ik\sum_{r=1}^{n}(N_{r}s_{r}-s_{r})}=(m_{21})_{n}^{*}\nonumber\\
\end{eqnarray}
where
\begin{subequations}
\begin{align}
(X_{r})_{n}&=U_{N_{r}-2}(\xi_{r})\prod_{p=r+1}^{n}(U_{N_{p}-1}(\xi_{p}))e^{ik\sum_{p=r}^{n}(N_{p}s_{p})}\nonumber\\
&\quad \times e^{-ik\sum_{p=r+1}^{n}(s_{p})}
\end{align}
\end{subequations}
and the argument $\xi_{n}$ is
\begin{align}
\xi_{n}&=(\text{Re}(m_{11})\cos(k\Lambda_{n})-\text{Im}(m_{11})\sin(k\Lambda_{n}))\nonumber\\
&\quad\times \prod_{r=1}^{n-1}(U_{N_{r}-1}(\xi_{r}))-\sum_{r=1}^{n-1}(\text{Re}((X_{r})_{n-1})\cos(ks_{n})\nonumber\\
&\quad-\text{Im}((X_{r})_{n-1})\sin(ks_{n}))-\cos(k(N_{n-1}s_{n-1}\nonumber\\
&\quad-s_{n})) U_{N_{{n}-1}-2}(\xi_{n-1})
\end{align}
where,
\begin{equation}
\Lambda_{n}=\sum_{r=1}^{n-1}(N_{r}s_{r}-s_{r})-s_{n}\label{}
\end{equation}
Then the transmission and reflection probability is,
\begin{eqnarray}
T(N_{1},N_{2},N_{3},\cdots ,N_{n})=\frac{1}{1+|(m_{12})_{n}|^{2}}\\
R(N_{1},N_{2},N_{3},\cdots ,N_{n})=\frac{|(m_{12})_{n}|^{2}}{1+|(m_{12})_{n}|^{2}}\label{21a}
\end{eqnarray}
The above equation can be simplified to
\begin{eqnarray}
T(N_{1},N_{2},N_{3},\cdots ,N_{n})=\frac{1}{1+[|m_{12}|\prod_{r=1}^{n}(U_{N_{r}-1}(\xi_{r}))]^{2}}\label{super periodic}\nonumber\\\\
R(N_{1},N_{2},N_{3},\cdots ,N_{n})=\frac{[|m_{12}|\prod_{r=1}^{n}(U_{N_{r}-1}(\xi_{r}))]^{2}}{1+[|m_{12}|\prod_{r=1}^{n}(U_{N_{r}-1}(\xi_{r}))]^{2}}\label{22a}\nonumber\\
\end{eqnarray}
\subsection{For two-dimensional scattering}
\begin{figure}[htb]
\centering
\includegraphics[width=\columnwidth]{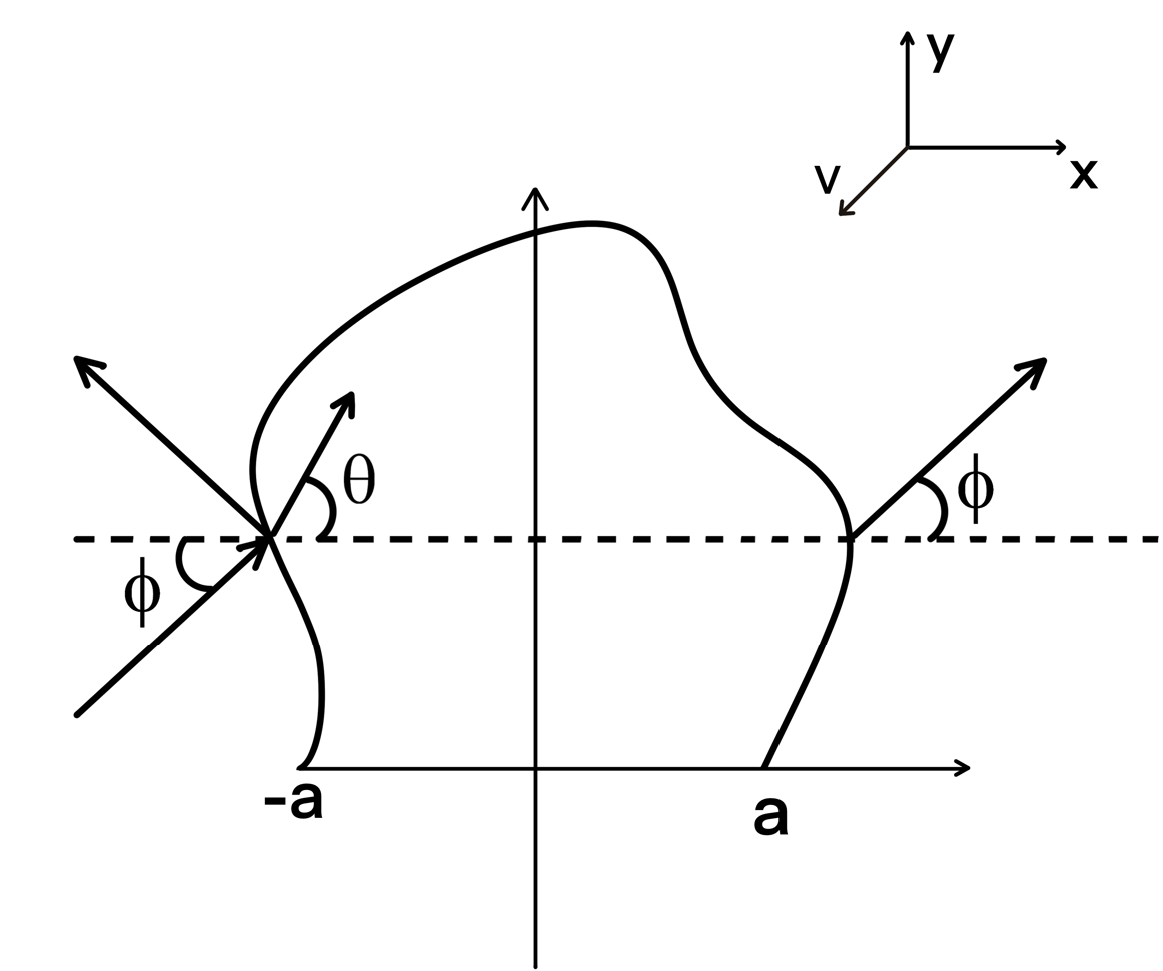}
\caption{Scattering from a arbitrary potential.}
\label{fig:13}
\end{figure}
Now, we discuss the transfer matrix method for studying a special case ($y$-component of the momentum is conserved) of two-dimensional scattering of particles from a potential, represented by a potential $V(x)$ for $-a < x < a$ as shown in FIG. \ref{fig:13}. Because of the conservation of the $y$-component of the momentum, the general form of the wave function can be written as
\begin{equation}
\psi(x,y)=\begin{cases}(Ae^{ik_{x}x}+Be^{-ik_{x}x})e^{ik_{y}y}~~~~~\text{ for  }x<-a\\
(\phi(x))e^{ik_{y}y}~~~~~~~~~~\text{ for  }-a<x<a\\
(Ce^{ik_{x}x}+De^{-ik_{x}x})e^{ik_{y}y}~~~~~\text{ for  }x>a\end{cases}
\end{equation}
Where $k$ is wave vector, $k_{x}=k\cos{\phi}$ and $k_{y}=k\sin{\phi}$ are the wave vector components outside the barrier.
By applying the appropriate boundary conditions at the endpoints $-a$ and $a$, which typically require the continuity of $\psi(x,y)$ and its derivative. Through this approach, we can establish two linear relationships among the coefficients $A$, $B$, $C$, and $D$ similar to the one-dimensional scenario. We can determine the transmission and reflection probabilities for super-periodicity by replacing $k$ with $k_{x}$ in the one-dimensional formulae. This adaptation is due to the potential repeating in the $x$-direction and the momentum is conserved in the $y$-direction.
\bibliography{References}
\end{document}